\documentclass[12pt]{article}

\usepackage{verbatim}
\usepackage{amsmath}
\usepackage{amssymb}
\usepackage{graphicx}
\usepackage{cite}
\usepackage{subfig}
\usepackage{setspace}
\usepackage{url}
\usepackage[percent]{overpic}
\usepackage{slashed}
\usepackage{authblk}

\usepackage{xspace}

\usepackage{fullpage}
\usepackage{hyperref}

\def\ie{{\it i.e.}}
\def\eg{{\it e.g.}}
\def\etc{{\it etc}}
\def\etal{{\it et al.}}

\def\to{\rightarrow}

\title{Fun with New Gauge Bosons at 100 TeV}
\date{}
\author{Thomas G. Rizzo}
\affil{SLAC National Accelerator Laboratory, Menlo Park, CA, 94025, USA\footnote{rizzo@slac.stanford.edu}}

\begin{document}

\rightline{\vbox{\halign{&#\hfil\cr
&SLAC-PUB-15917\cr
}}}


{\let\newpage\relax\maketitle}

\begin{abstract}
The production of new gauge bosons is a standard benchmark for the exploration of the physics capabilities of future colliders. The 
$\sqrt s=100$ TeV Future Hadron Collider will make a major step in our ability to search for and explore the properties of such new 
states. In this paper, employing traditional models to make contact with the past and more recent literature, we not only establish 
in detail the discovery and exclusion reaches for both the $Z'$ and $W'$ within these models but, more importantly, we also examine 
the capability of the FHC to extract information relevant for the determination of the couplings of the $Z'$ to the fermions of the 
Standard Model as well as the helicity of the corresponding $W'$ couplings. This is a necessary first step in determining the nature 
of the underlying theory which gave rise to these states.
\end{abstract}

\section{Introduction and Background}

Interest is growing in the physics of a possible Future Hadron Collider (FHC) which will take over the Energy Frontier sometime after the 
running of the HL-LHC been has completed{\cite {FHC}}. CERN has already begun a 5-year study to investigate this possibility{\cite {CERN}}. 
Presently, the FHC is currently envisioned as having a center-of-mass energy of $\sqrt s \simeq 100$ TeV and having the capability to accumulate  
integrated luminosities of order $\sim 1-10$ ab$^{-1}$ or greater{\footnote {Interestingly, this energy represents as large of a step above 
that of the LHC as the LHC was above the Tevatron.}}. As is well-known, the LHC has/had `the origin of electroweak symmetry breaking in 
the Standard Model' as a known 'no-loose' physics target to set its energy scale. Clearly, with the discovery of the Higgs boson in 2012, 
the LHC has already been quite successful in this regard. The 100 TeV energy for the FHC, on the otherhand, represents an educated extrapolation 
of what might be possible given foreseeable technological improvements without incurring `prohibitive' costs while simultaneously allowing for 
the opening up of potential new physics thresholds. The FHC is thus truly a machine of exploration as were the accelerators of earlier generations. 
Of course, in a more general context, the physics associated with such higher energy hadron colliders beyond the LHC has been frequently discussed 
from time to time over the past two decades{\cite {VLHC}} since the cancellation of the SSC. 

One of the historic and standard new physics benchmarks that is always employed in the study of the potential capabilities of future colliders 
is the production and examination of new neutral ($Z'$) and charged ($W'$) gauge bosons{\cite {rev}} the reason being that they are a 
relatively common occurrence in many beyond the Standard Model (BSM) scenarios and they (usually) have relatively clean leptonic signatures. 
This may be 
of particular relevance at a high luminosity $\sqrt s=100$ TeV machine where one can presently only guess at the difficulty of making measurements  
within this challenging hadronic environment. While previous works{\cite {zpfuture}} have provided us with some idea as to what to expect 
as far as the discovery potential for $Z',W'$ might be at the FHC, here we wish to go somewhat further in both the quantification of 
these results and to address the question of how well such new states might be studied assuming that they are indeed discovered. In 
order to make direct contact with many past studies{\cite {rev,zpfuture,Han:2013mra}} we will restrict ourselves to the somewhat traditional 
set of BSM models containing $Z'/W'$ considered there. We remind the reader that this does not in any way exhaust the range of possibilities 
for such new states but it does give us a respectable range of predictions to examine for what one might expect at 100 TeV. Here we will 
also restrict our attention to the relatively safe final states which employ at least one (very) high $p_t$ lepton trigger due to the 
unknown nature of the hadronic environment and pile-up conditions the detectors may have to deal with at 100 TeV as mentioned earlier. 
Furthermore, we will concentrate mostly on electron final states as muon $p_t$ resolution for (many) mult-TeV muons is likely to be difficult 
with the typical magnetic fields that are currently available given finite detector size even when we allow for factors of two scaling in 
both these quantities. We note that if the muon pair mass resolution is sufficiently degraded this will have a significant impact on narrow $Z'$ 
searches in this channel.

Present measurements by both ATLAS{\cite {ATLAS:2013jma,Aad:2011fe}} and CMS{\cite {CMS:2013qca,CMS:2013rca}} at $\sqrt s=8$ TeV with 
$\sim 20$~fb$^{-1}$ of integrated luminosity can only provide lower bounds on the masses of possible new $Z',W'$ states. For example, if 
these states have couplings identical to those of the corresponding SM gauge bosons (\ie, the so-called SSM `model') their masses must 
exceed 2.96 and 3.35 TeV, respectively. Of course for other types of couplings, quite different, generally somewhat weaker results are 
obtained as in the case of an $E_6$-inspired $Z'${\cite {it}}. Eventually, at the 14 TeV HL-LHC, search sensitivities for the same $Z_{SSM}'$ 
and $W_{SSM}'$ states may be as large as $\sim 6-8$ TeV{\cite {HL-LHC}} depending upon whether they are discovered or excluded. In what 
follows we will generally assume that the masses of these new states are beyond the reach of the HL-LHC at least to study in any detail even 
if they were to be in fact discovered there.

In the analysis below we will examine not only the discovery and exclusion rates for these new states at 100 TeV but also survey how data 
obtainable by experiments at the FHC may be used to determine their detailed properties through a number of different observables. This survey 
is not meant to be either exhaustive or conclusive but only to provide a first glimpse at what some of these possibilities might be based on 
earlier studies performed for other hadron colliders{\footnote {We will not consider the interesting possibility to study this new physics 
with polarized proton beams here\cite{Fuks:2014uka}.}}. Much work will still be needed to examine the feasibility of the use of the many 
observables discussed below for the study of new gauge bosons at the FHC. We first consider new $Z'$ gauge bosons in the next Section and 
then turn our attention to the $W'$ case in Section 3. Section 4 contains a discussion and our conclusions. The Appendix considers the case of 
SSM $Z'$ and $W'$ gauge bosons but with non-SSM coupling strengths.

\section{New $Z'$ Bosons}

The first issue we address is the obvious one: what is the mass reach for $Z'$ gauge boson discovery or exclusion at $\sqrt s=100$ TeV? The usual 
approach as performed 
at the LHC has two components: ($i$) compare the expected opposite-sign dilepton invariant mass distribution with the SM expectations and look for 
excesses at high mass and then either claim an excess or set a limit ($ii$) compare the expected number of (excess?) dilepton events with the the 
narrow-width approximation (NWA) cross 
section (times leptonic branching fraction) for the the $Z'$ within a given BSM model. To perform the corresponding calculations here for the 
FHC we will assume a detector below with a lepton rapidity coverage ($|\eta_\ell|\leq 2.5$) and electron EM energy resolution similar 
to that of the present ATLAS detector and, to be specific, we will make use of the default NLO CTEQ6.6 pdfs{\cite {Nadolsky:2008zw}} and scale 
the relevant cross sections employing NLO K-factors{\cite{Melnikov:2006kv}} when obtaining our numerical results{\footnote {We remind the 
reader that there are presently significant uncertainties in extrapolating the currently available pdfs up to 100 TeV. We note that employing 
the NNLO CT10 pdfs gives the same cross section results as used here up to corrections of order $5\%$ percent.}}. $Z'$ (and correspondingly 
$W'$ below) partial widths will be calculated including NLO and partial NNLO QCD corrections as well as the corresponding NLO QED corrections.
As a final comment, we note that these calculations will be performed using a fixed-width prescription for the $Z'$ but one can easily check that 
identical results are obtained using the running-width instead given the resulting level of statistics.

\begin{figure}[htbp]
\centerline{\includegraphics[width=5.0in,angle=90]{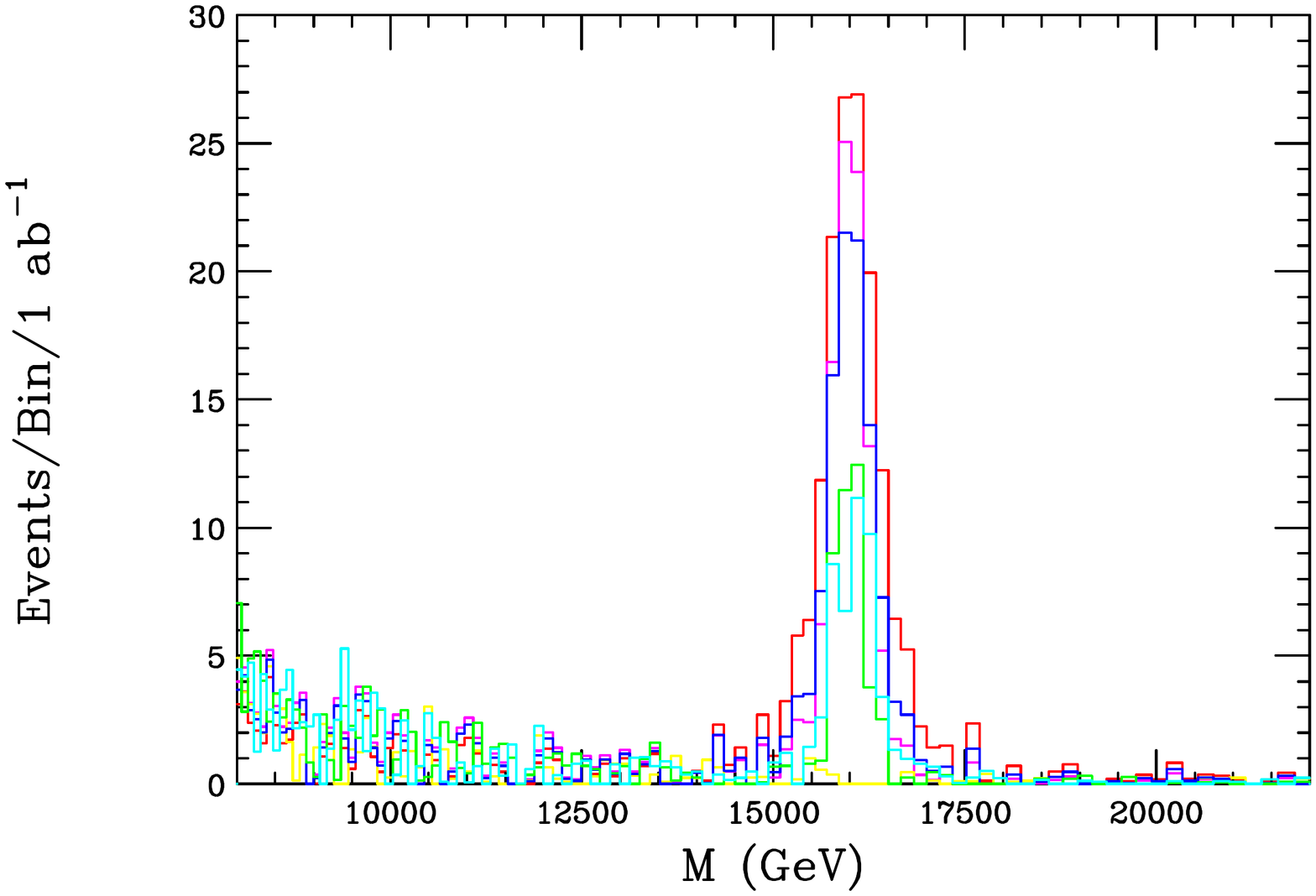}}
\vspace*{-4.0cm}
\centerline{\includegraphics[width=5.0in,angle=90]{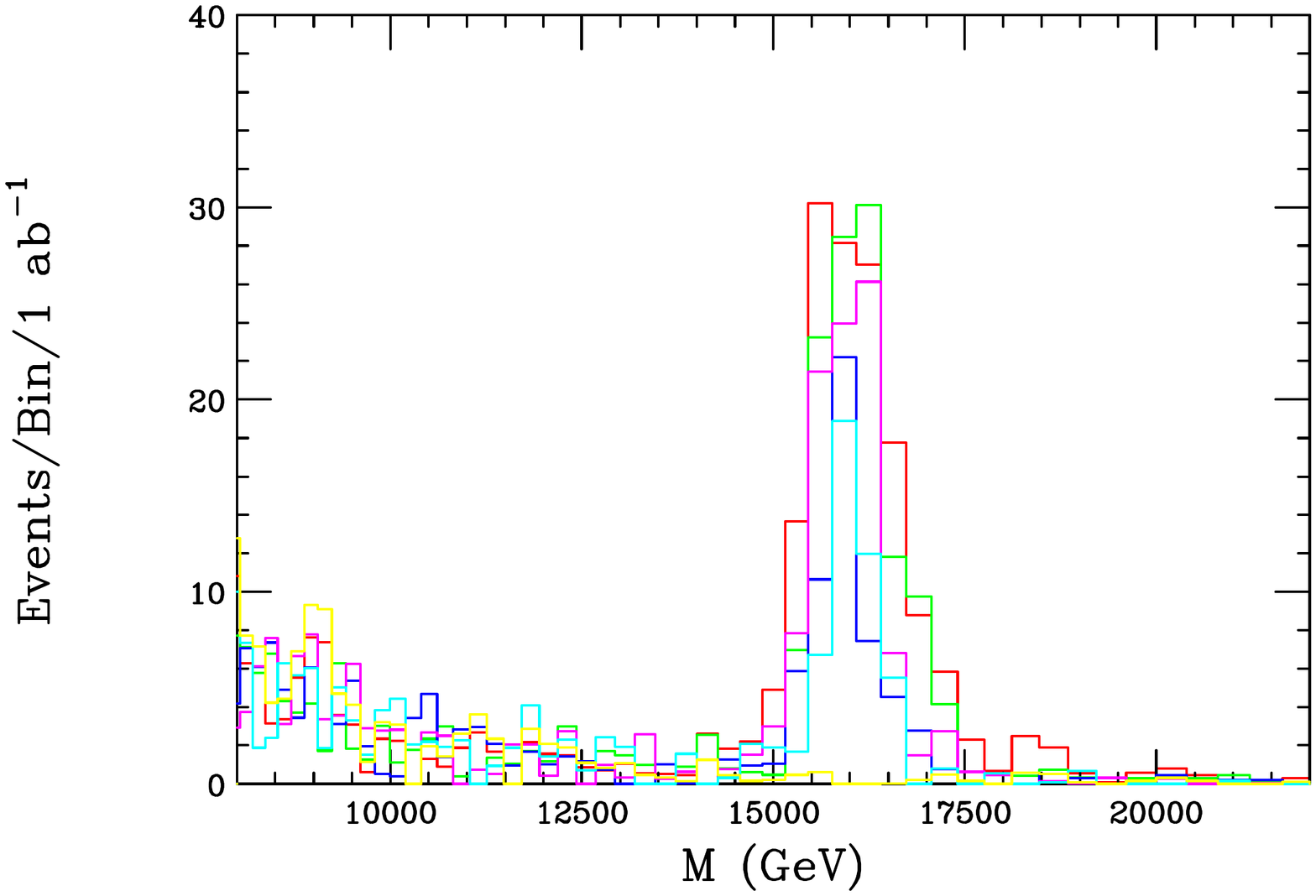}}
\vspace*{-1.90cm}
\caption{Top: Histograms of the Drell-Yan mass distribution for the production of a 16 TeV $Z'$ at the FHC: the red (green, blue, magenta, cyan) 
histogram is for the SSM (Left-Right Model (LRM)~\cite{LRM}, $E_6$ models $\psi,~\chi$ and $\eta$, respectively). The yellow histogram is 
the SM background. ATLAS-type acceptances and smearings have been assumed. Bottom: Same as the upper panel but now assuming a factor of 2 worse 
dielectron mass resolution; note the change in the size of the binning here.}
\label{fig01}
\end{figure}

\begin{figure}[htbp]
\centerline{\includegraphics[width=5.0in,angle=90]{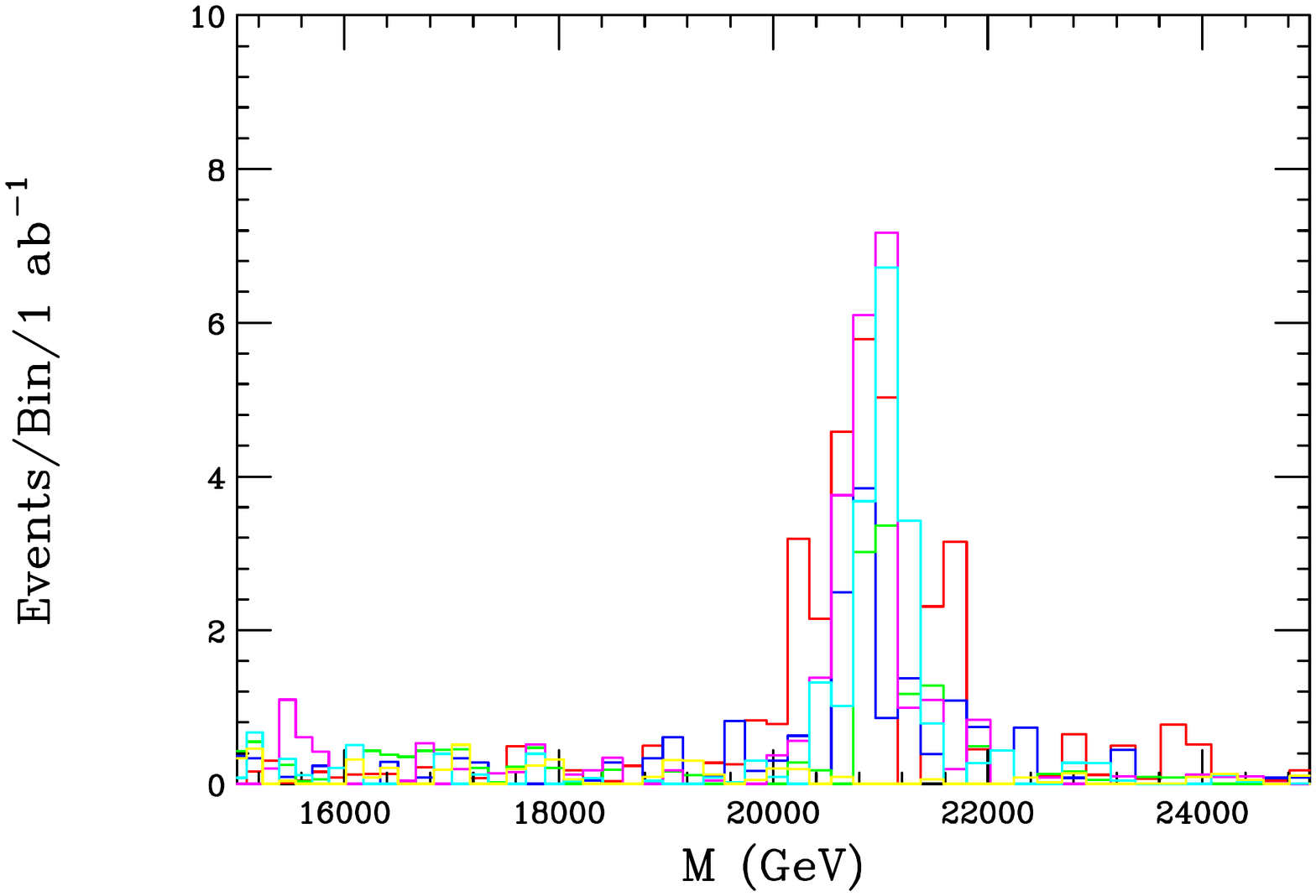}}
\vspace*{-4.0cm}
\centerline{\includegraphics[width=5.0in,angle=90]{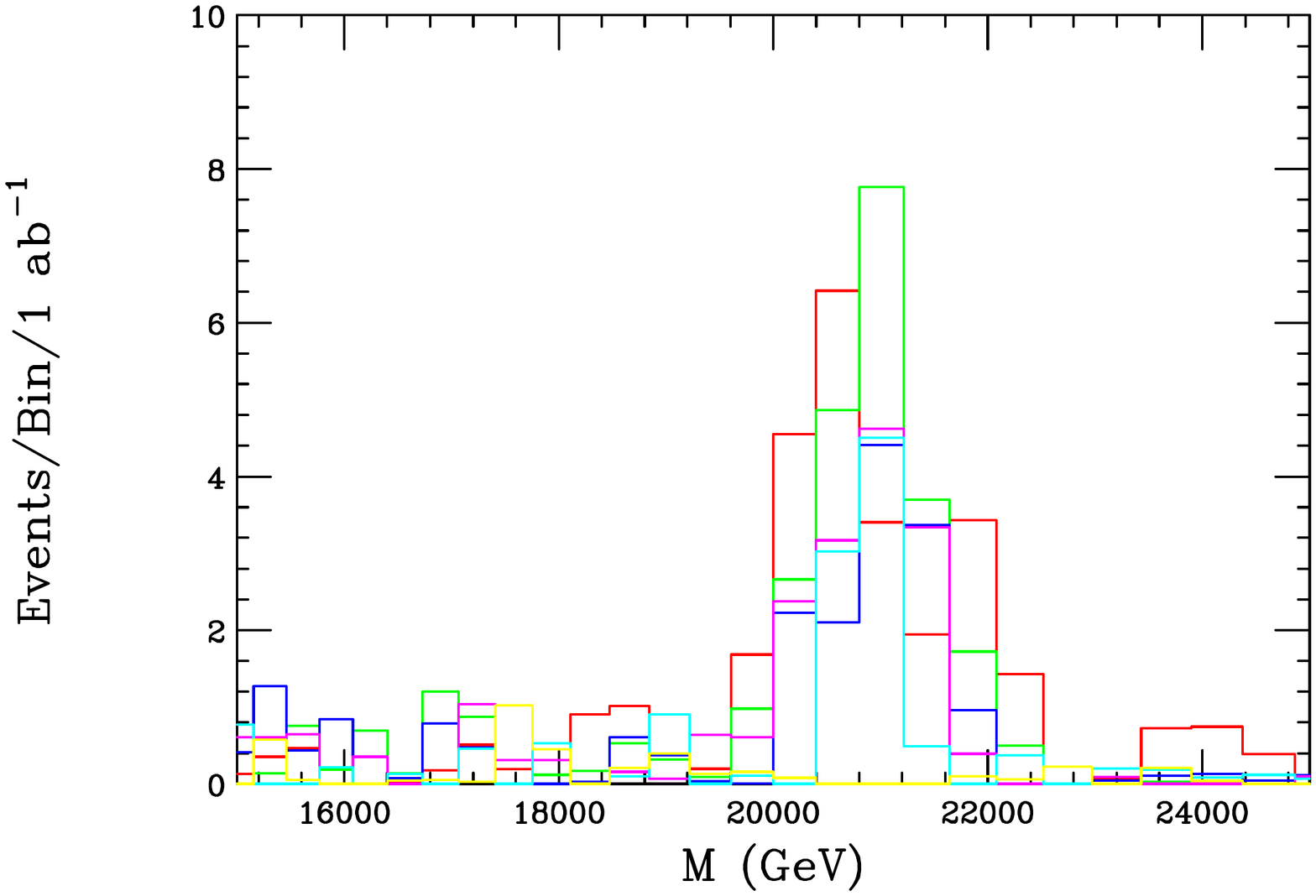}}
\vspace*{-1.90cm}
\caption{Same as the previous Figure but now for a $Z'$ mass of 21 TeV.}
\label{fig02}
\end{figure}

To begin our analysis it is instructive to examine the resonance signal structure for $Z'$ production in the Drell-Yan dielectron mass 
distribution at 100 TeV. A first example of this is shown in Figs.~\ref{fig01} and ~\ref{fig02} which show the Drell-Yan mass distributions for 
several different $Z'$ models assuming a mass of either 16 or 21 TeV along with the associated SM backgrounds taking an integrated luminosity of 
1 ab$^{-1}$.{\footnote {In what follows we will always assume that any new gauge bosons are un-mixed with their SM counterparts and that they 
decay only into the known SM states in obtaining all numerical results.}} From the upper panel of Fig,~\ref{fig01}, it is fairly obvious that 
for a mass of 16 TeV with this luminosity all of these models predict an easily visible signal above the SM background. In the lower panel we 
see how the signals will respond if the dilepton mass resolution were to become worse by a factor of 2; clearly all the states still 
remain easily visible. 
It is important to note that in most of these models the width to mass ratio of the $Z'$ is quite comparable to and frequently less than these 
resolutions which imply that it is the resolution itself that often controls the `lineshape' and not the actual $Z'$ width.  
On the otherhand, when the $Z'$ mass is increased to 21 TeV (not a randomly chosen number), as seen in Fig.~\ref{fig02}, it is a bit less 
obvious how statistically convincing the signal remains in all these different cases. However, by counting events, we find that most of these 
models {\it would} likely be discovered although several lie (very) close to the statistical boundary $\sim 10$ signal events in this mass region 
since it essentially background-free. Of course, the corresponding dimuon sample could also be potentially helpful here but given the likely 
inferior mass resolution for dimuons its not clear by how much without some further assumptions. Certainly the dimuon distribution would also 
likely show a high mass excess confirming that in dielectrons but would not yield a very reliable determination of the peak mass value since this 
excess would likely be quite broadly distributed. It would seem that for this class of models $\simeq 20-21$ TeV represents a rough estimate of 
the rough lower mass limit for discovery.

\begin{figure}[htbp]
\centerline{\includegraphics[width=5.0in,angle=90]{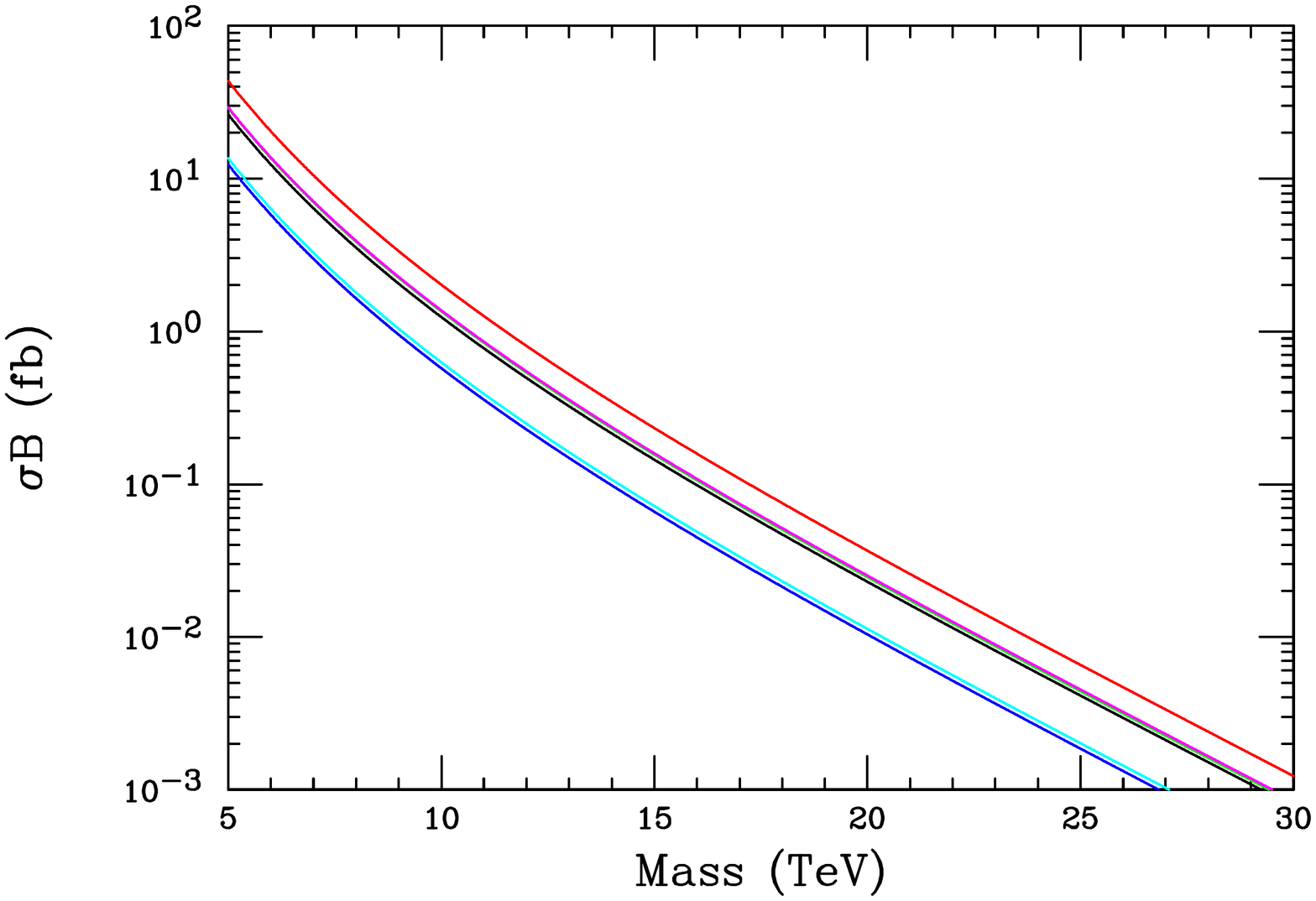}}
\vspace*{-4.0cm}
\centerline{\includegraphics[width=5.0in,angle=90]{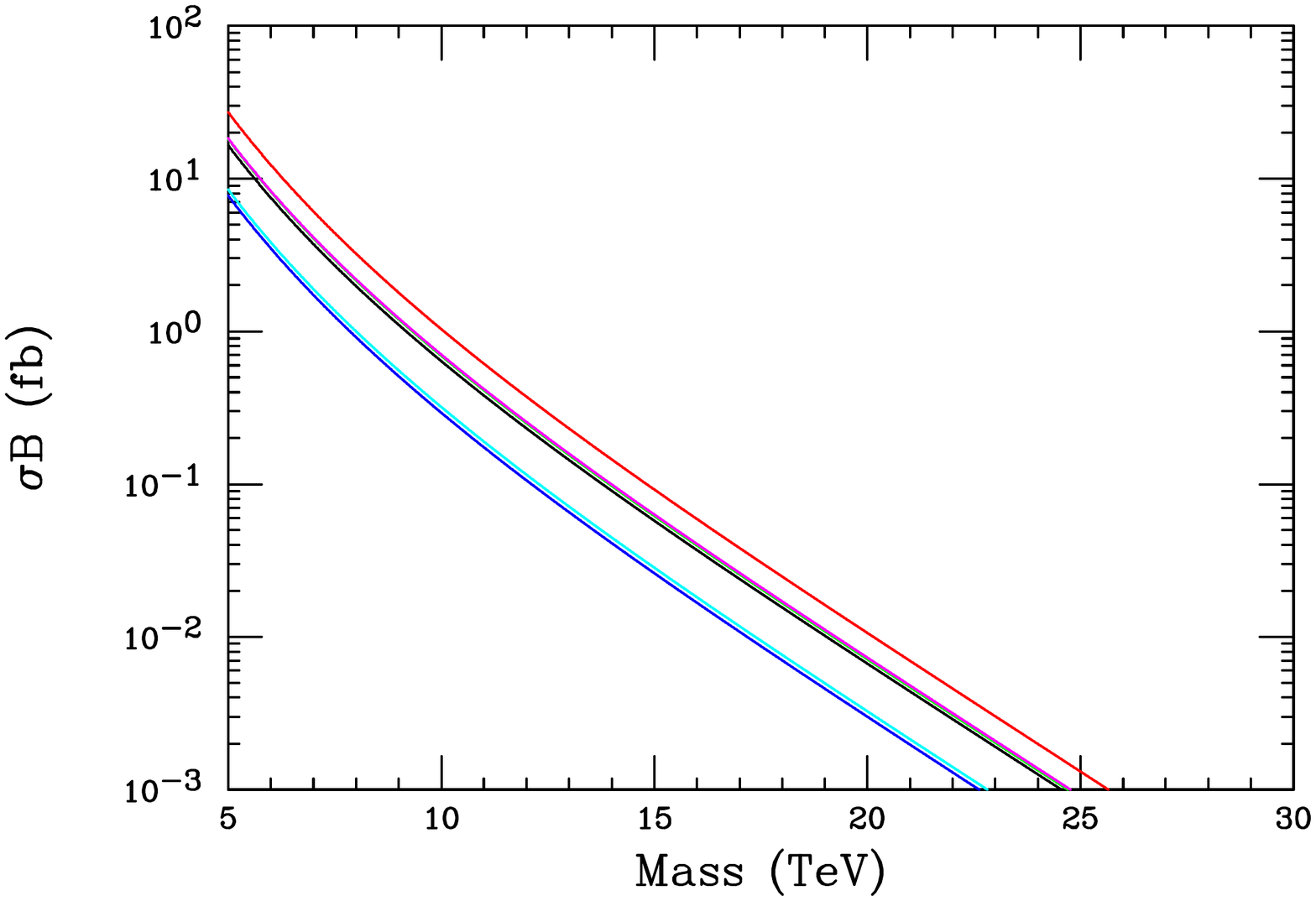}}
\vspace*{-1.50cm}
\caption{$\sigma B_\ell$ in fb as a function of the $Z'$ mass in TeV at $\sqrt s=100(80)$ TeV calculated in the NWA in the top(bottom)
panel. The color codes for the curves shown here are the same as in the previous Figure but now also include (black solid) that for 
$E_6$ model I.}
\label{fig03}
\end{figure}

These results on both potential discovery and possible exclusion can be further summarized by the content of Fig.~\ref{fig03} and in 
Table~\ref{rates1}. Fig.~\ref{fig03} shows the dilepton production signal cross section via the $Z'$ arising in different models 
calculated in the NWA including acceptance cuts for both $\sqrt s=80$ and 100 TeV. In this Figure we see several things: for example, 
even within this rather restricted set of models 
the cross sections can vary by about a factor of $\simeq 5$ at any given mass value. Furthermore, we see that at large masses, for a 
fixed value of the cross section, going from $\sqrt s=80$ to 100 TeV will allow us to increase the reach by roughly $\simeq 4.2$ TeV. 
For later discussions, 
Table~\ref{rates1} shows the actual numerical values of these cross sections for $Z'$ masses of 10, 15 and 20 TeV and also 
the expected discovery and exclusion reaches in TeV (assuming an integrated luminosity of 1 ab$^{-1}$) for the same set of models. These 
values will be important when we examine statistical errors on the various quantities we introduce below. We also note here that an  
increase in the integrated luminosity by a factor of 3(10) will increase both the discovery and exclusion reaches by roughly 
$\simeq 2.9(5.7)$ TeV for all these models.

\begin{table}
\centering
\begin{tabular}{|l|c|c|c|c|c|} \hline\hline
Model    & 10 TeV  & 15 TeV  & 20 TeV  &   Disc. & Excl.   \\
\hline

SSM        & 2021.  & 232.6 &  36.65 & 23.8 & 27.3  \\
LRM        & 1353.  & 156.1 &  24.62 & 22.6 & 26.1  \\
$\psi$     & 573.7  & 65.93 &  10.37 & 20.1 & 23.6  \\
$\chi$     & 1372.  & 159.0 &  25.18 & 22.7 & 26.2  \\
$\eta$     & 626.8  & 71.82 &  11.38 & 20.3 & 23.8  \\
I          & 1241.  & 144.4 &  22.94 & 22.4 & 25.7  \\
\hline\hline
\end{tabular}
\caption{$\sigma B_\ell$ in ab at 100 TeV for different $Z'$ model masses are shown in the left three columns employing the NWA. Discovery and 
exclusion reaches in TeV for 1 ab$^{-1}$ of integrated luminosity are visible in the two right-hand columns. Increasing the integrated luminosity 
to 3 ab$^{-1}$ will raise the reach mass values in all cases by $\simeq 2.9$ TeV.}
\label{rates1}
\end{table}

Once a $Z'$ has been discovered, we want to learn {\it which} model, if any, it corresponds to which means that we need to learn about 
its couplings to the various SM fields. If these couplings are generation independent (which would appear likely given current flavor 
physics constraints) this will involve 7 independent parameters. Furthermore, if the new gauge group to which the $Z'$ belongs commutes 
with the SM $SU(2)_L$ group factor, which is true for GUT groups such as the LRM and $E_6$, there remain only 5 such parameters 
corresponding to the couplings of the $Z'$ to $u_L=d_L$, $\nu_L=e_L$, and $(u,d,e)_R${\footnote {Of course the SSM does {\it not} fall in this 
class as it is merely used as a benchmark so that there are still 7 distinct couplings in this case as for the SM $Z$ (by definition).}}.  
Note that within these GUT-like frameworks any gauge anomalies that may apparently exist are canceled by the presence of additional 
fermions which are either SM singlets or are vector-like with respect to the SM interactions. As noted above, when making the numerical 
calculations that we will present below it will always be assumed that all such fields are kinematically inaccessible in the decays of the 
new gauge bosons that we consider. However, since such states {\it might} participate in $Z', W'$ decays, using observables that are 
sensitive to the existence of these states in order to identify the $Z'$ or $W'$ themselves is {\it problematic} and may lead to the wrong 
conclusions\cite {rev}. Interestingly, we note that reducing the leptonic branching fraction by a factor of two due to new open channels 
will result in a reduction of the discovery and exclusion reaches by roughly $\simeq 2$ TeV in all cases. 

It should be noted here for completeness that the observation of a new high mass resonance in the opposite-sign dilepton channel does not 
necessarily imply that a $Z'$ has been discovered. For example, we have not discussed here the importance of the determination of the 
resonance's spin via the leptonic angular distribution~\cite {rev}. One could imagine that such a resonance could be spin-2 (\eg, a Kaluza-Klein 
graviton) or even spin-0 (\eg, an R-Parity violating sneutrino) so the discovery of a spin-1 state is assumed in the discussion that follows.   

\begin{figure}[htbp]
\centerline{\includegraphics[width=5.0in,angle=90]{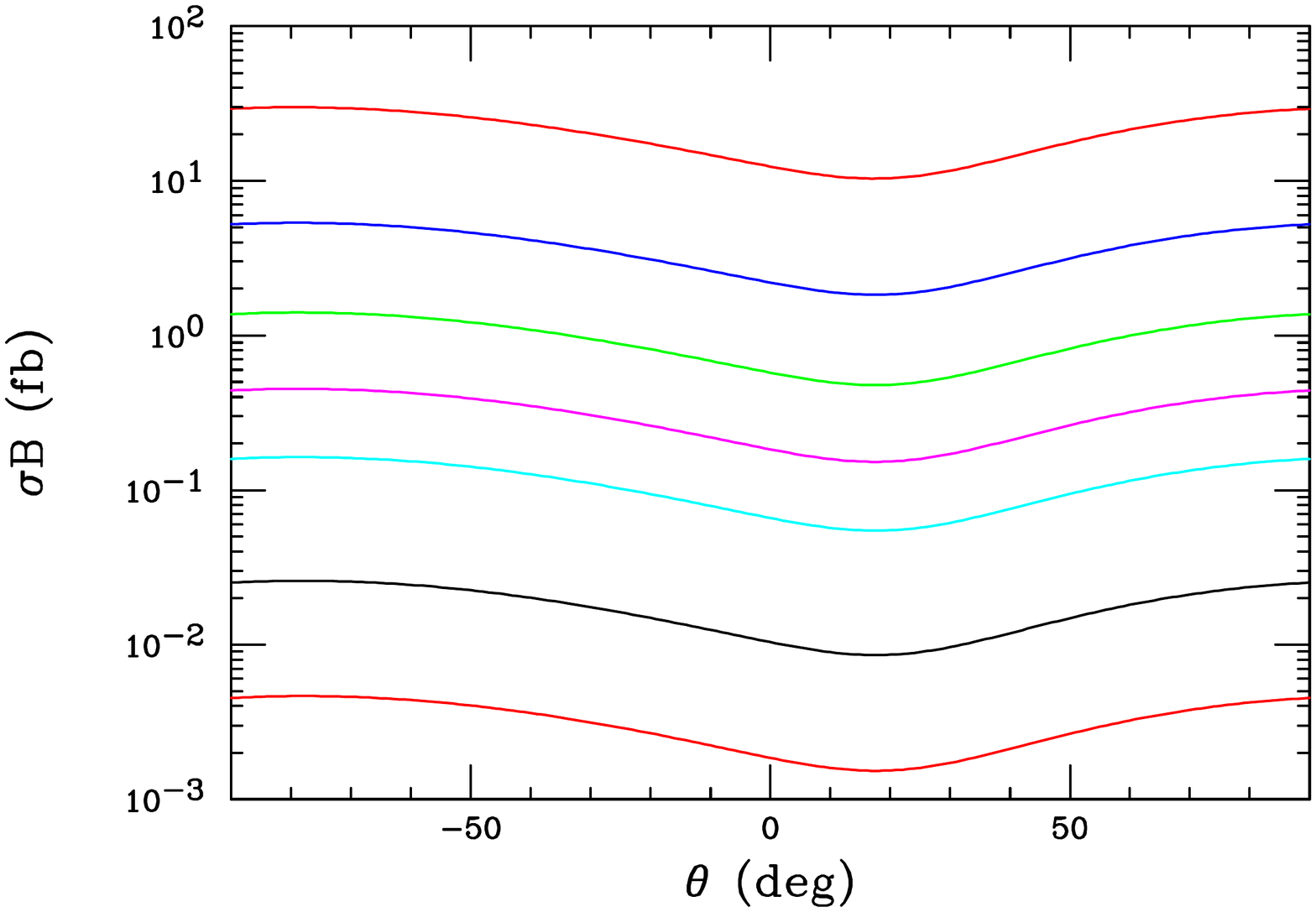}}
\vspace*{-4.0cm}
\centerline{\includegraphics[width=5.0in,angle=90]{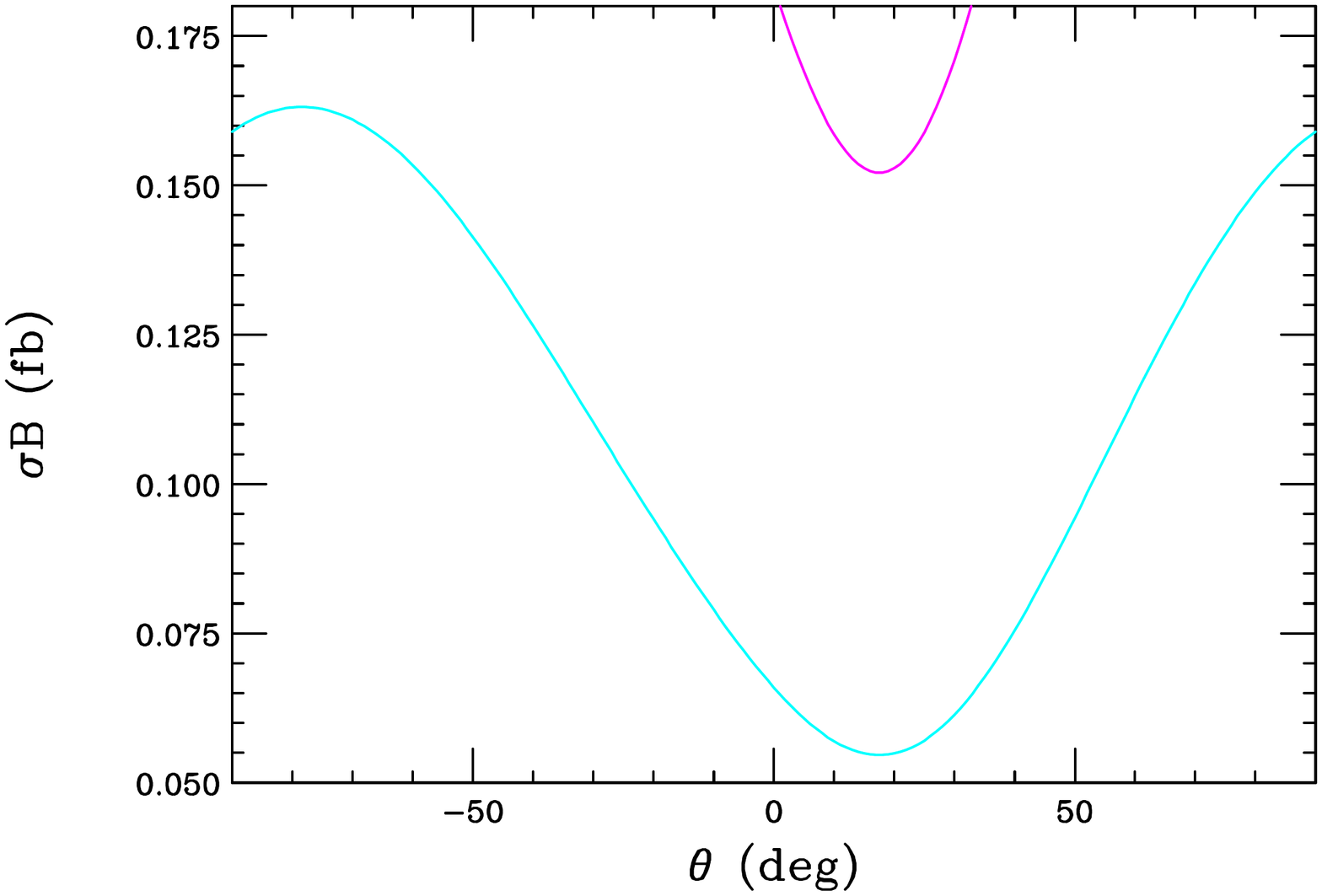}}
\vspace*{-1.50cm}
\caption{$\sigma B_\ell$ in fb as a function on the $E_6$ mixing angle $\theta$. In the top panel we see the result for masses (from top 
to bottom) of 5, 7.5, 10, 12.5, 15, 20 and 25 TeV, respectively. In the lower panel we see the magnified result for the case of M=15 TeV.}  
\label{fig04}
\end{figure}

Clearly the production cross section times leptonic branching fraction of the $Z'$ itself is a potentially interesting observable for 
determining fermionic couplings. As we saw in Table~\ref{rates1} above, these cross sections vary substantially from model to model, 
even within the $E_6$ scenario itself as is shown in Fig.~\ref{fig04} where a variation of more than a factor of 3 is observed. Trivially, 
this observable on its own is obviously sensitive to variations in the different coupling parameters. However, the value of the leptonic branching 
fraction employed to obtain these results assumes that the $Z'$ can decay to only SM particles. Thus, strictly speaking, we cannot use 
this observable (alone) to obtain determinations of the $Z'$ couplings. Of course, as has long been known, {\it ratios} of such cross section 
times branching fractions are insensitive to any additional decay modes that the $Z'$ may have and we will make use of this result below.

\begin{figure}[htbp]
\centerline{\includegraphics[width=5.0in,angle=90]{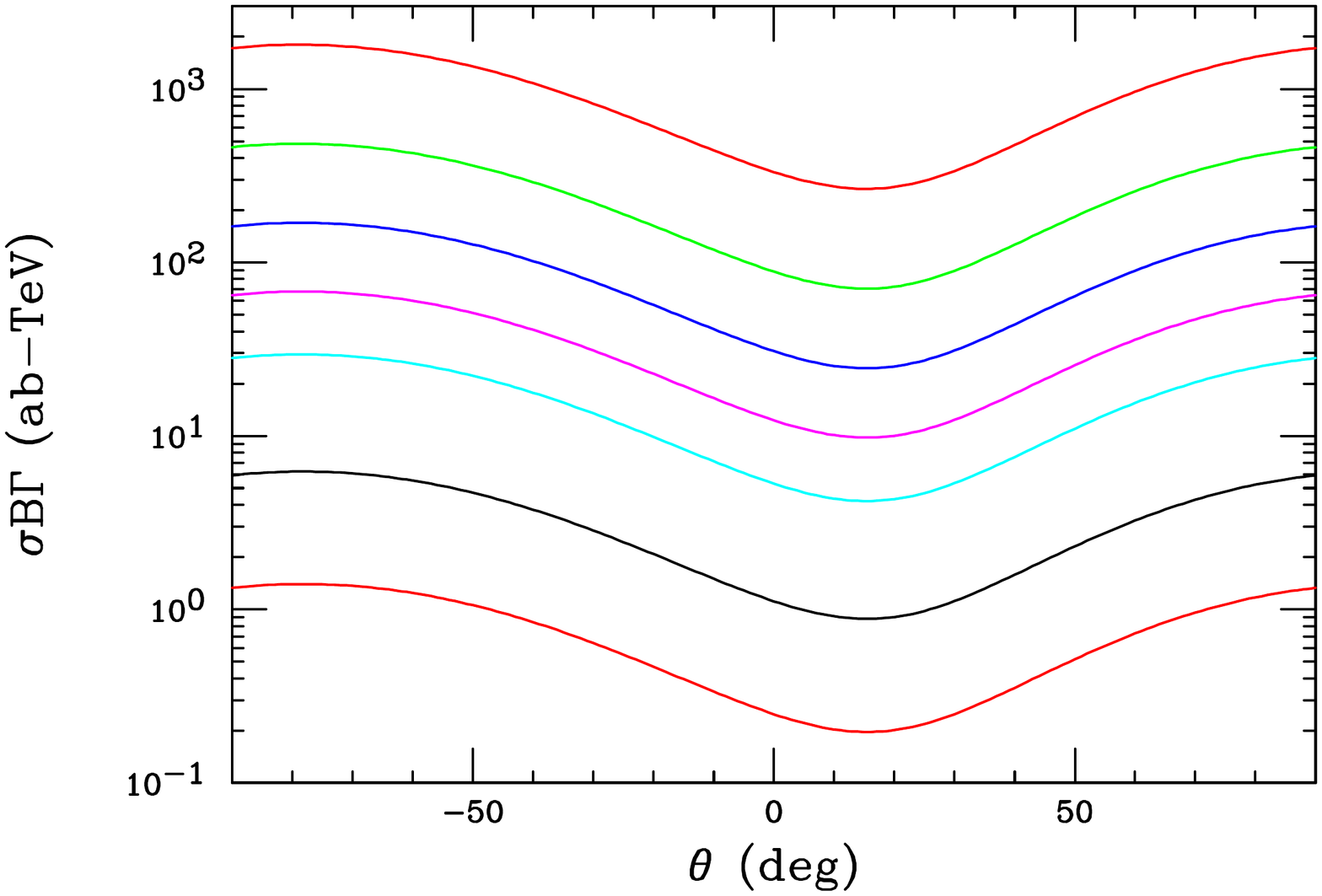}}
\vspace*{-4.0cm}
\centerline{\includegraphics[width=5.0in,angle=90]{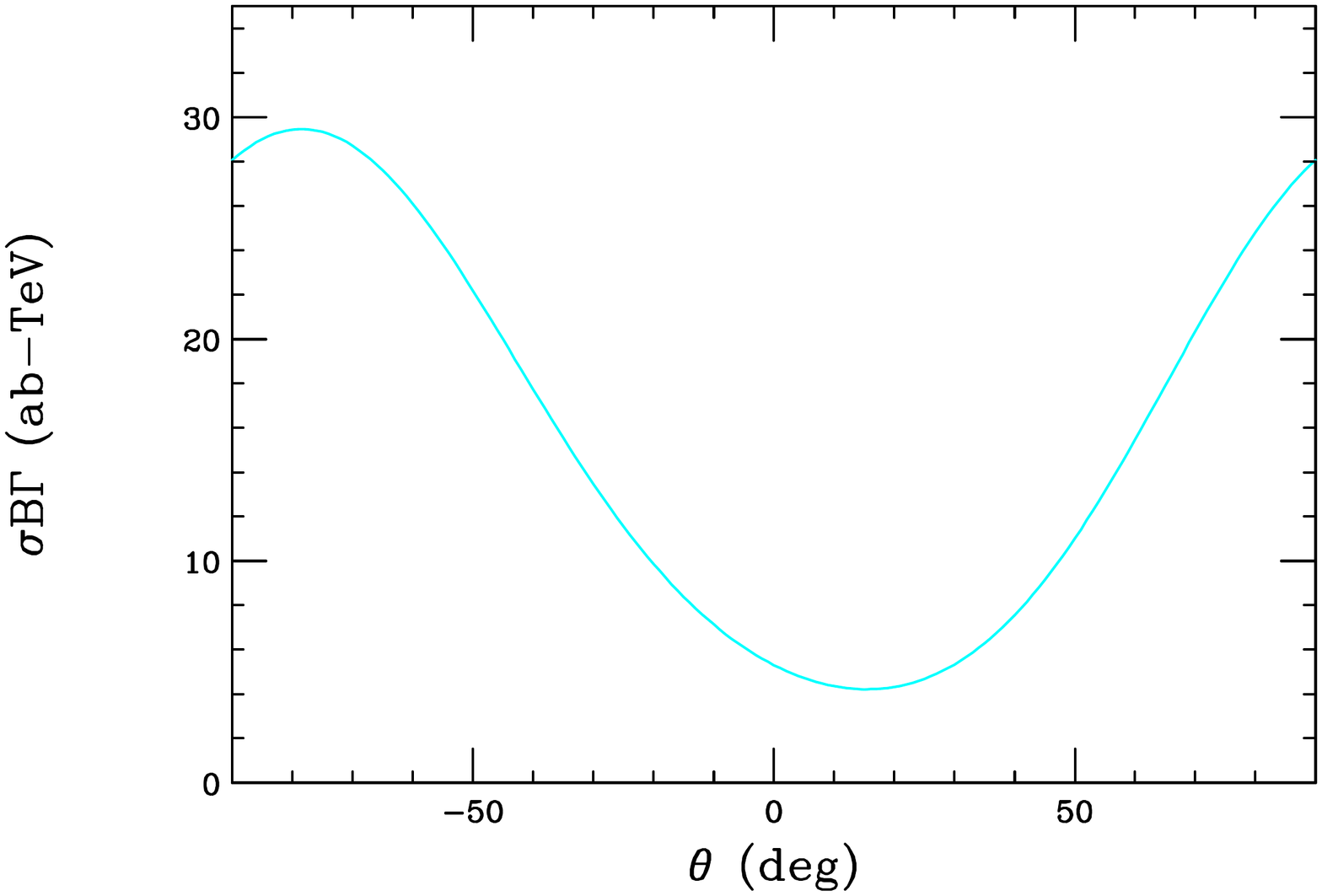}}
\vspace*{-1.50cm}
\caption{$\sigma B_\ell \Gamma_{Z'}$ at 100 TeV in units of ab-GeV as a function on the $E_6$ mixing angle $\theta$. The curves are labeled as in 
the previous Figure.}  
\label{fig04-5}
\end{figure}

With this issue in mind and using the same production channel (dileptons) as for discovery there are other observables that one can use to 
obtain coupling information that are insensitive to any potential additional $Z'$ decay models. The most obvious one is obtained by just rescaling 
the overall signal rate by the extracted width of the $Z'$ resonance. In the NWA we see this quantity is just the product 
$\sigma B_\ell \Gamma_{Z'}$ which is clearly independent of whether or not additional $Z'$ decay modes are present (since the product 
$B_\ell \Gamma_{Z'}$ is just the leptonic width of the $Z'$) and this remains true 
to a good approximation even when a more detailed calculation is performed. Results for this quantity using the NWA can be seen in 
Table~\ref{rates1-5} and in Fig.~\ref{fig04-5}. This Figure shows the NWA evaluation of this quantity in the 
$E_6$ case demonstrating its respectable coupling sensitivity, varying by a factor of $\simeq 6$ just within this model framework while the 
Table shows variations as large as a factor of $\simeq 15$. To use 
this quantity we need to extract not only the production cross section in the leptonic channel but also we need to determine the $Z'$ width 
by a de-convolution of the mass resolution in the peak region. An early ATLAS study of this quantity for the 14 TeV LHC\cite{Schafer} for 
10 fb$^{-1}$ of simulated data suggests that this may be possible with slightly larger than statistical errors, \eg,  $\simeq 1.3/\sqrt N$. 
Clearly further study of this quantity at higher energies and masses would be valuable.

\begin{table}
\centering
\begin{tabular}{|l|c|c|c|} \hline\hline
Model    & 10 TeV  & 15 TeV  & 20 TeV  \\
\hline

SSM        & 0.6108  & 0.1054 &  0.0221 \\
LRM        & 0.2815  & 0.0487 &  0.0102 \\
$\psi$     & 0.0308  & 0.0053 &  0.0011 \\
$\chi$     & 0.1615  & 0.0281 &  0.0059 \\
$\eta$     & 0.0404  & 0.0069 &  0.0015 \\
I          & 0.1326  & 0.0231 &  0.0049 \\
\hline\hline
\end{tabular}
\caption{$\sigma B_\ell \Gamma_{Z'}$ in units of fb-TeV at $\sqrt s=100$ TeV for different $Z'$ model masses employing the NWA.}
\label{rates1-5}
\end{table}

\begin{figure}[htbp]
\centerline{\includegraphics[width=5.0in,angle=90]{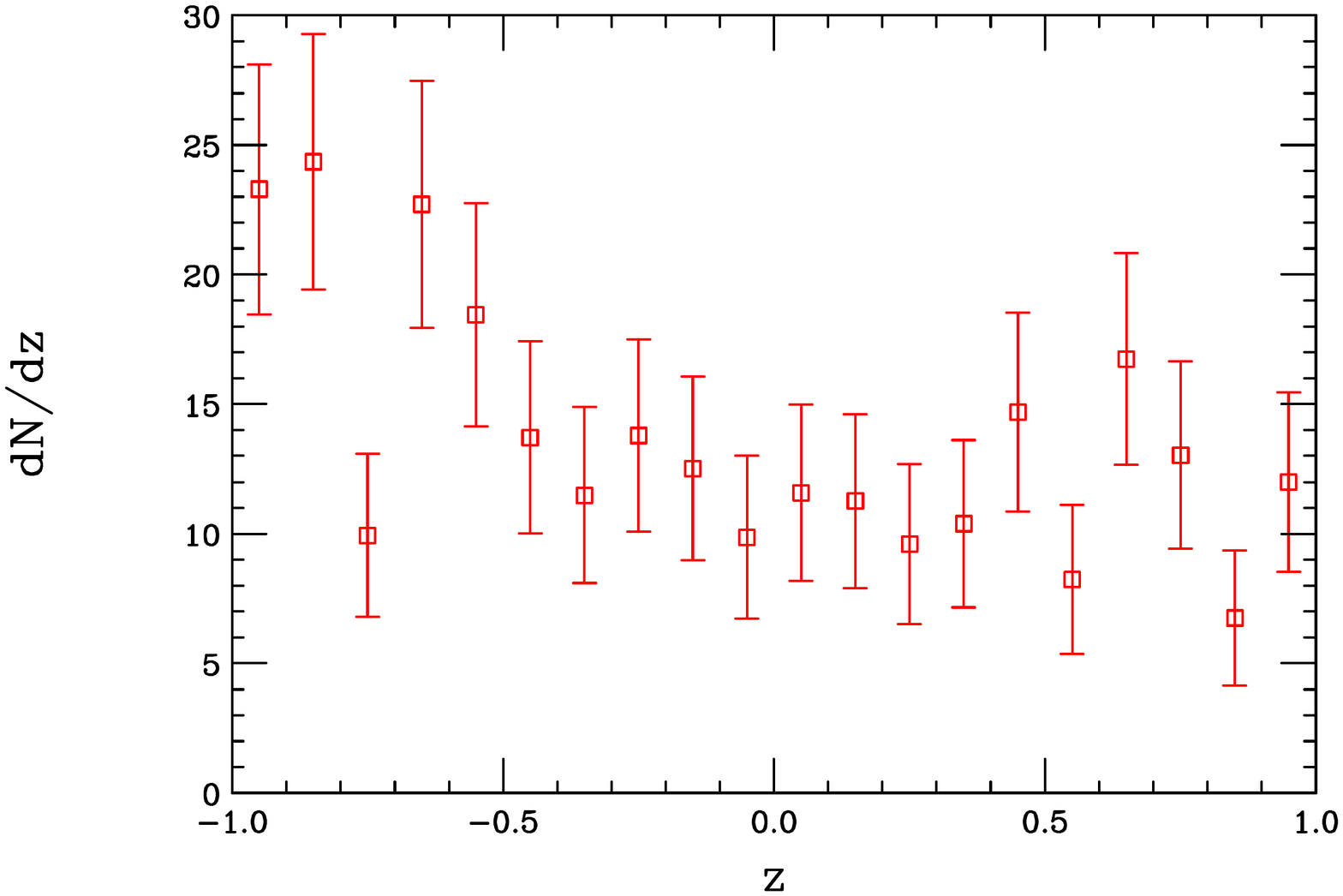}}
\vspace*{-4.0cm}
\centerline{\includegraphics[width=5.0in,angle=90]{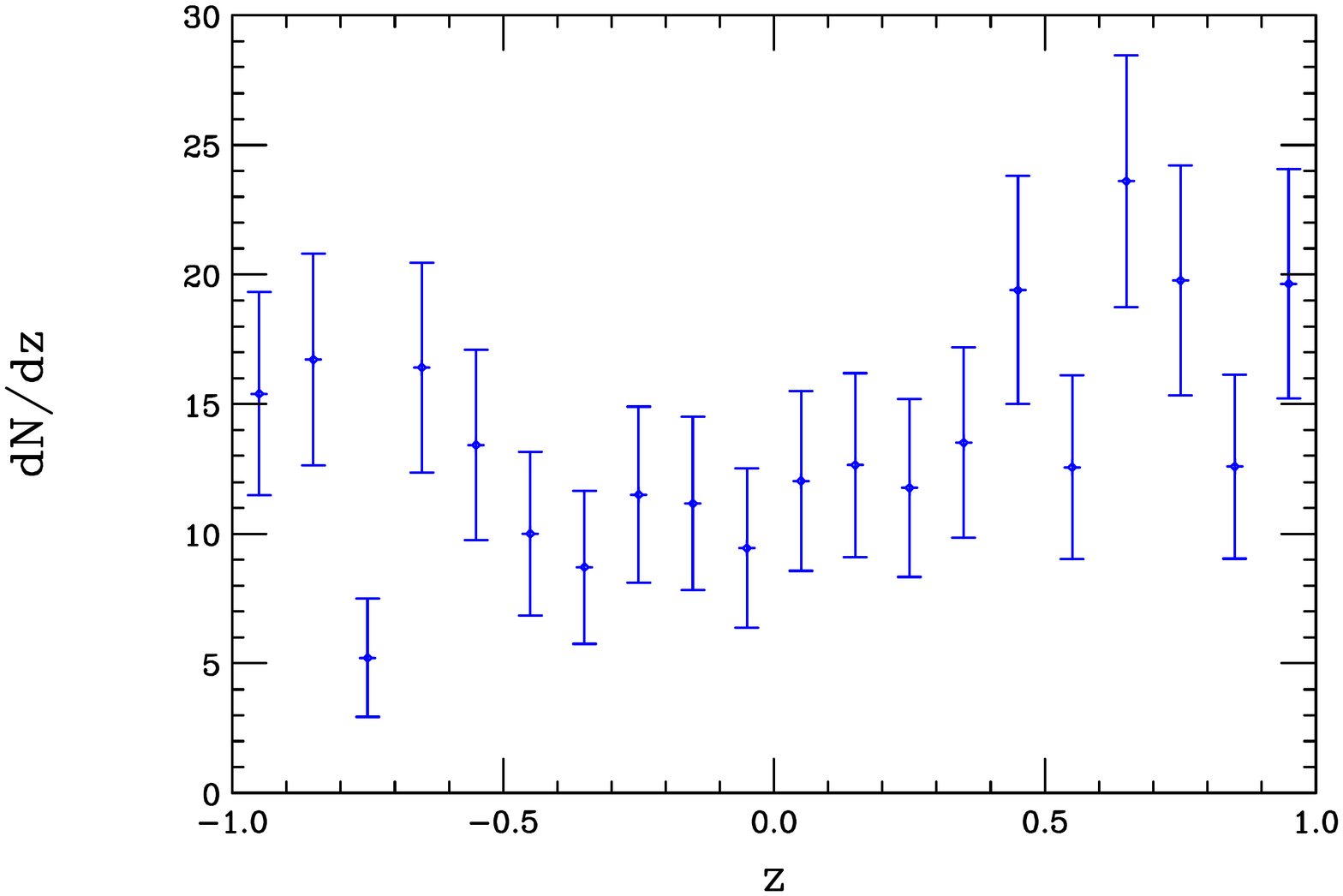}}
\vspace*{-1.50cm}
\caption{Idealized normalized leptonic angular distributions assuming 300 observed events assuming $A_{FB}=-0.2(0.1)$ in the top(bottom) panel. 
Only statistical errors are show.}  
\label{fig05}
\end{figure}

Perhaps the most well known example of a coupling-sensitive observable that makes use of the dilepton discovery channel is $A_{FB}$ which can 
be obtained in principle 
from the lepton's angular distribution. In the $Z'$ rest frame, with $z=\cos \theta^*$ being defined between the initial quark, $q$, and 
outgoing $l^-$ direction this distribution is given by $\sim 1+z^2 +8 A_{FB} z/3$. Fig.~\ref{fig05} shows two very idealized examples for these 
distributions (before any acceptance corrections, \etc ) assuming 300 dilepton events are observed. These figures demonstrate the typical 
level of statistics required to determine $A_{FB}$ with some reliability (given the typical asymmetries predicted in many models) in an idealized 
situation. To get closer to reality two additional requirements are obvious: ($i$) an acceptance correction needs to be applied to account for the 
cut of $|\eta_\ell|<2.5$ on the lepton rapidity. This effectively reduces the the number of events at large values of $|z|$ thus reducing the 
overall sensitivity to any non-zero asymmetry. 
Even more importantly, ($ii$) a correction must be applied to identify which direction along the collision axis is to be identified with 
that of the initial $q$. A first approximation for this (that can be later more fully performed by Monte Carlo) is to note that, {\it most 
of the time}, due to the pdfs, the $q$ direction can be identified with boost direction of the $Z'$ which can be easily reconstructed. 
This requires a {\it minimum} cut on $|y_{Z'}|$, here chosen to be $>0.3$ for numerical purposes, which, at the very least will further 
reduce the level of statistics. This cut becomes quite significant as the $Z'$ mass increases and the $Z'$ rapidity becomes ever more central 
with the maximum $Z'$ boost being $y_{max}=\rm log(\sqrt s/M_{Z'}) \simeq 2.30(1.90,1.61)$ for masses of 10(15,~20) TeV at $\sqrt s=100$ TeV.

This discussion informs us that a reasonable level of statistics, roughly on the order of $\sim 300$ signal events, will be required for 
a respectable determination of $A_{FB}$. Such event rates are achievable for $Z'$ masses roughly $\simeq 6$ TeV {\it below} the mass where 
discovery is possible. This implies, \eg, that for an SSM $Z'$ a reliable value of $A_{FB}$ is not obtainable if the mass is in excess of 18 
TeV or so unless luminosities in excess of the canonical 1 ab$^{-1}$ value that we have assumed here are achievable. 

\begin{figure}[htbp]
\centerline{\includegraphics[width=5.0in,angle=90]{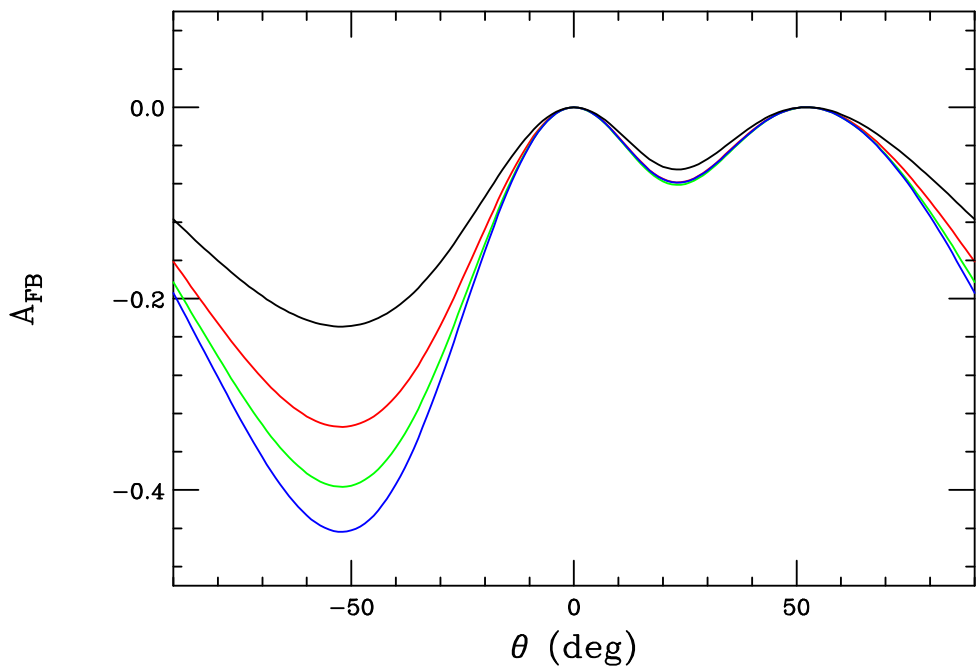}}
\vspace*{-0.90cm}
\caption{NWA values of $A_{FB}$ in $E_6$ models as above for $Z'$ masses of 5, 10, 15 and 20 TeV (from top to bottom) after the 
rapidity cuts described in the text have been applied.}  
\label{fig06}
\end{figure}

\begin{figure}[htbp]
\centerline{\includegraphics[width=5.0in,angle=90]{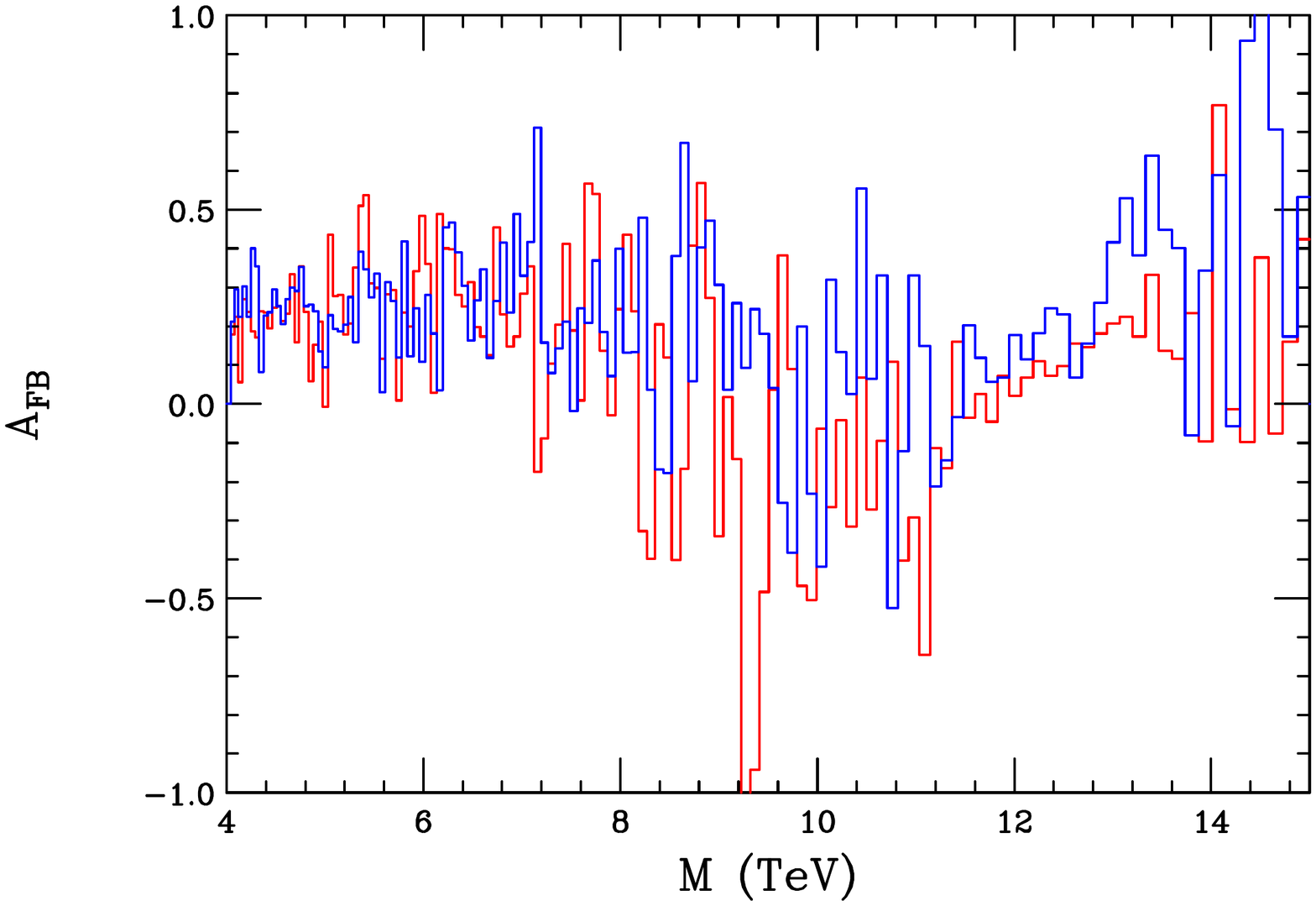}}
\vspace*{-4.0cm}
\centerline{\includegraphics[width=5.0in,angle=90]{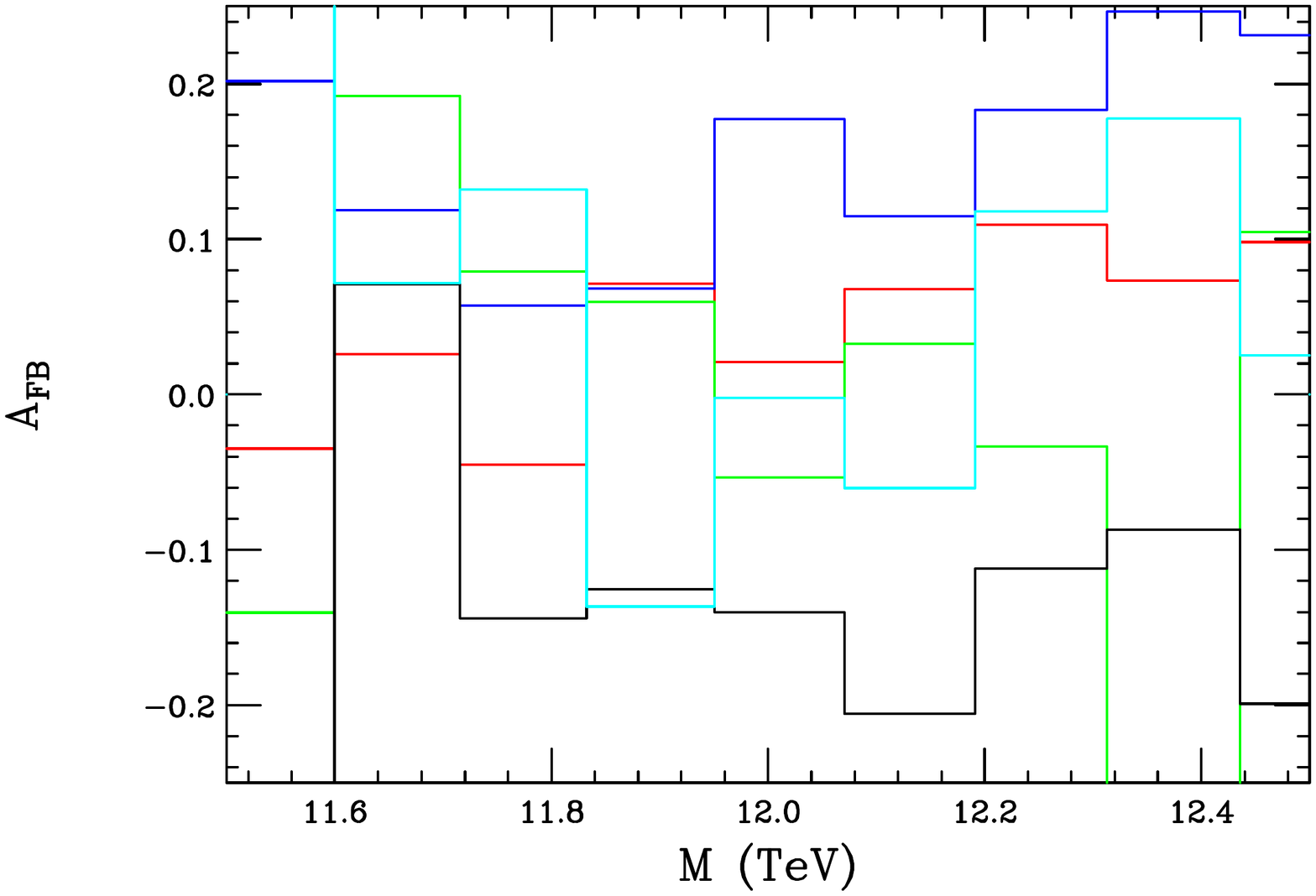}}
\vspace*{-1.50cm}
\caption{Dilepton invariant mass dependence of $A_{FB}$, subject to the cuts described in the text assuming a $Z'$ mass of 12 TeV and an 
integrated luminosity of 5 ab$^{1}$, (Top) For a range of narrow invariant mass bins for the SSM(red) and LRM(blue) scenarios. (Bottom) For the 
SSM(red), LRM(blue), $\psi$(green), $\chi$(black) and $\eta$(cyan) models but now tightly focused on the invariant mass region surrounding the 
12 TeV resonance region.}  
\label{fig07}
\end{figure}

\begin{figure}[htbp]
\centerline{\includegraphics[width=5.0in,angle=90]{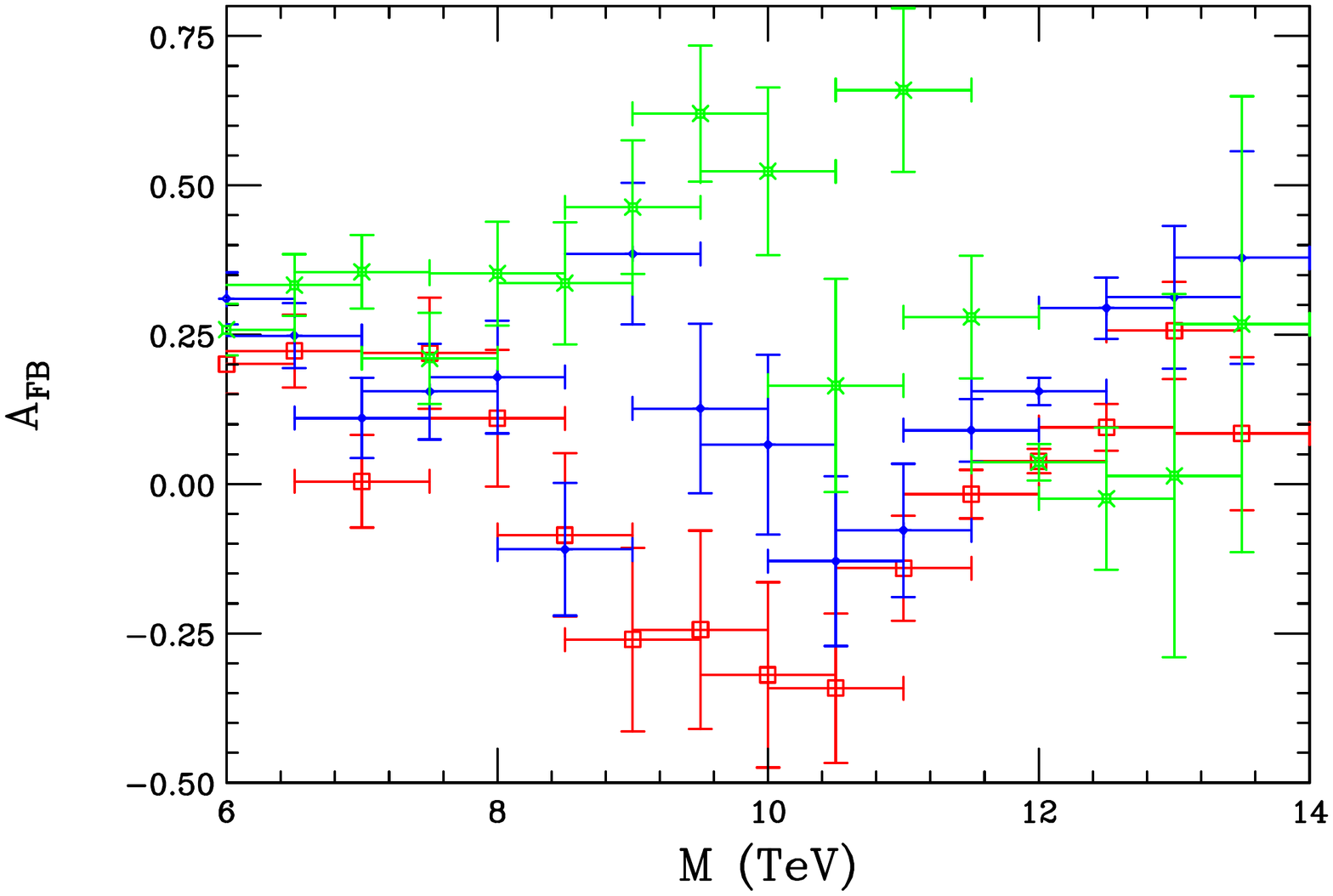}}
\vspace*{-4.0cm}
\centerline{\includegraphics[width=5.0in,angle=90]{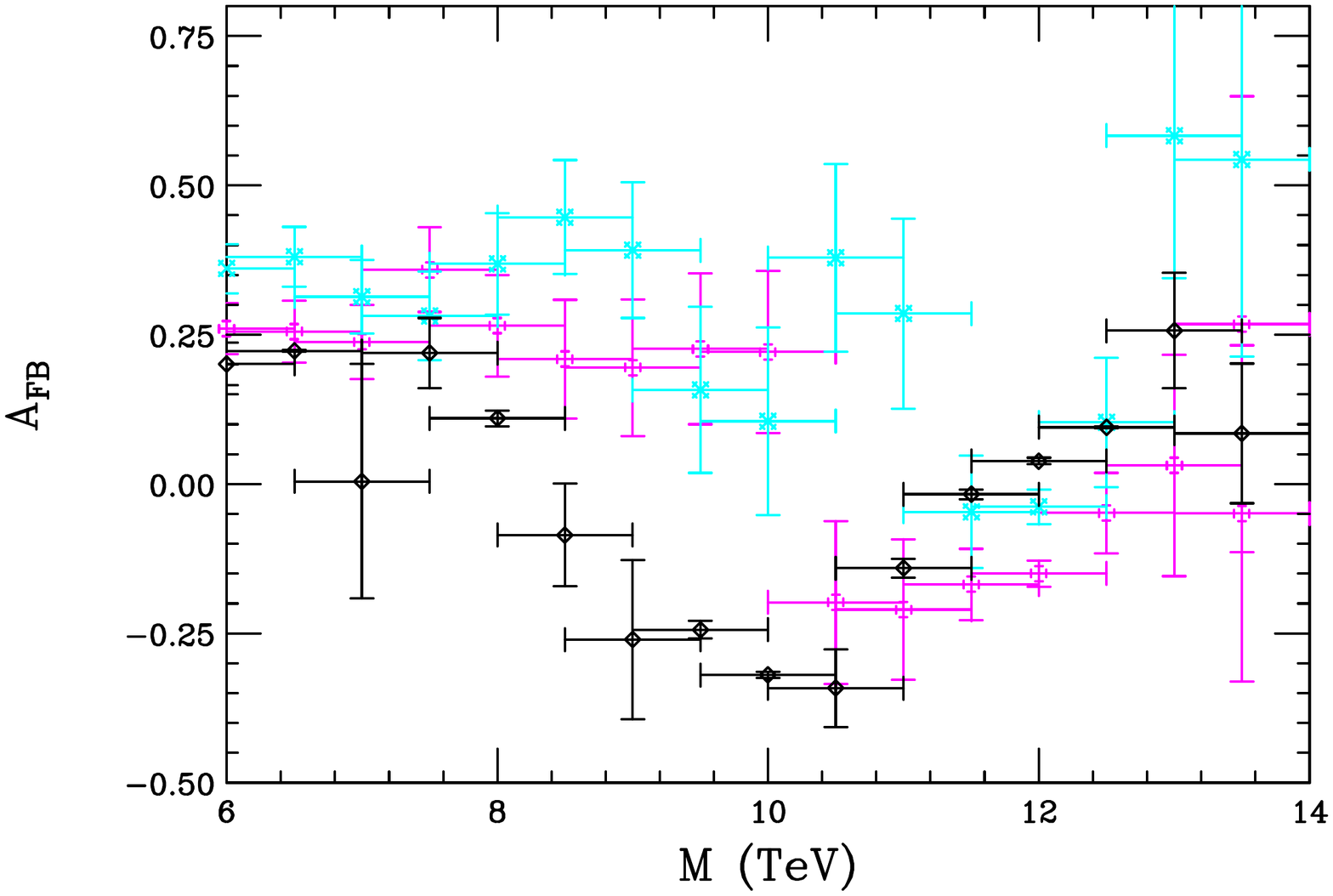}}
\vspace*{-1.50cm}
\caption{Dilepton invariant mass dependence of $A_{FB}$, subject to the cuts described in the text assuming a $Z'$ mass of 12 TeV and an 
integrated luminosity of 5 ab$^{1}$ integrated over 500 GeV wide mass bins to gain statistics. These are shown for (Top) the SSM(red), LRM(blue), 
$\psi$(green), (Bottom) $\chi$(magenta), $\eta$(cyan) and I models.}  
\label{fig071}
\end{figure}

Of course, if $A_{FB}$ is indeed measurable with some precision it will provide the SM fermion coupling sensitivity as advertised as can 
be seen in the upper panel of Fig~\ref{fig06} and in Table~\ref{rates2} both of which were obtained in the NWA. Here we see not only the 
model dependence of $A_{FB}$ but also its dependence on the $Z'$ mass due to the evolution of the pdfs as well as the rapidity requirements 
due to the kinematically induced changes in the the corresponding 
distributions. A possible advantage of going beyond the NWA is that $A_{FB}$ as a function of the dilepton mass can be examined providing 
additional information about the interference between the SM and $Z'$ exchanges{\cite {rev}}.  Fig.~\ref{fig07} shows that in a more 
realistic situation 
where the NWA is not employed, getting a good handle on $A_{FB}$ in narrow mass bins (outside of the one with the resonance peak itself) 
will be difficult even for $Z'$ masses as low as 12 TeV and even with luminosities of 5 ab$^{-1}$ as shown here. The bin to bin fluctuations 
are seem to be far dominant here due to the limited statistics. Thus only the coarse-grained invariant mass dependence of $A_{FB}$ will likely 
be of any use away from the pole. Fig.~\ref{fig071} shows that this is indeed the case, making use of fixed 500 GeV wide bins over which 
$A_{FB}$ is averaged.  
Here we clearly see  that measurements of $A_{FB}$ in the SM-$Z'$ interference regime will be useful in coupling extraction provided 
sufficient luminosity is available to reduce the statistical errors. We note that in the lower panel of Fig.~\ref{fig07}, which examines just the 
bins surrounding and containing the $Z'$ mass peak, that the NWA provides a reasonable estimate of the actual asymmetry near the pole. Note that 
once we have passed the pole statistics plummets and we gain little more information.

\begin{table}
\centering
\begin{tabular}{|l|c|c|c|c|} \hline\hline
Model    & 5  TeV   & 10 TeV  &  15 TeV  & 20 TeV   \\
\hline

SSM        & 0.0363  & 0.0472 &  0.0491 & 0.0485  \\
LRM        & 0.0795  & 0.1056 &  0.1113 & 0.1115  \\
$\psi$     & 0       &  0     &  0      &  0     \\
$\chi$     & -0.1171  & -0.1612 &  -0.1827 & -0.1942  \\
$\eta$     & -0.0274  & -0.0334 &  -0.0347 & -0.0340  \\
I          & -0.2292  & -0.3339 &  -0.3966 & -0.4437  \\
\hline\hline

\end{tabular}
\caption{$A_{FB}$ at $\sqrt s=100$ TeV for different $Z'$ model masses employing the NWA and the rapidity cuts as described in the text.}
\label{rates2}
\end{table}

\begin{figure}[htbp]
\centerline{\includegraphics[width=5.0in,angle=90]{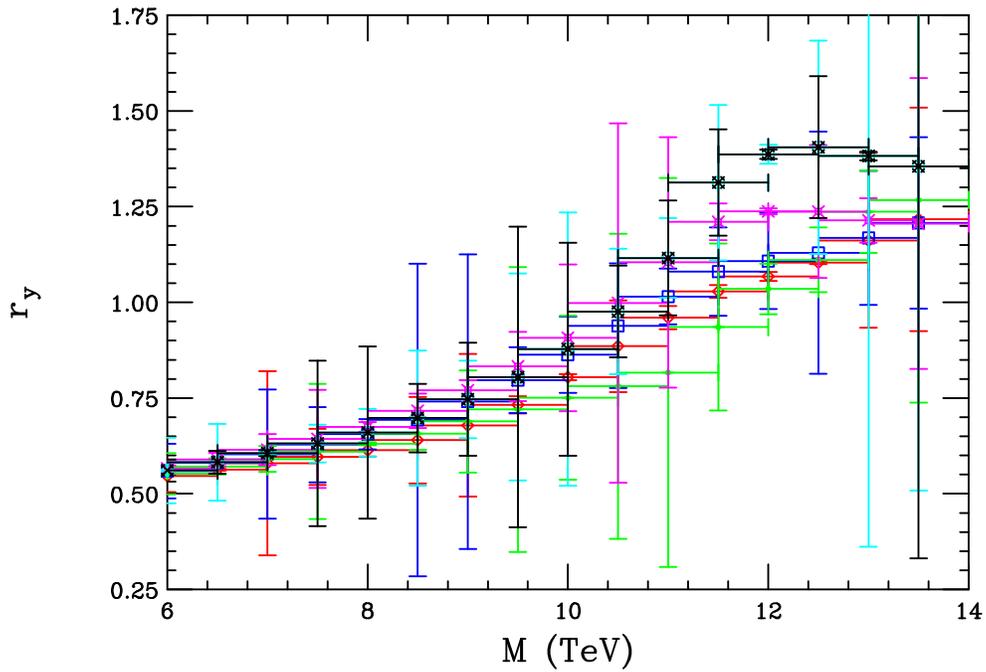}}
\vspace*{-0.90cm}
\caption{The invariant mass dependence of $r_y$, subject to the cuts described in the text assuming a $Z'$ mass of 12 TeV and an 
integrated luminosity of 5 ab$^{1}$ integrated over 500 GeV wide mass bins. These are shown for the SSM(red), LRM(blue), 
$\psi$(green), $\chi$(magenta), $\eta$(cyan) and I models.}  
\label{fig072}
\end{figure}

\begin{figure}[htbp]
\centerline{\includegraphics[width=5.0in,angle=90]{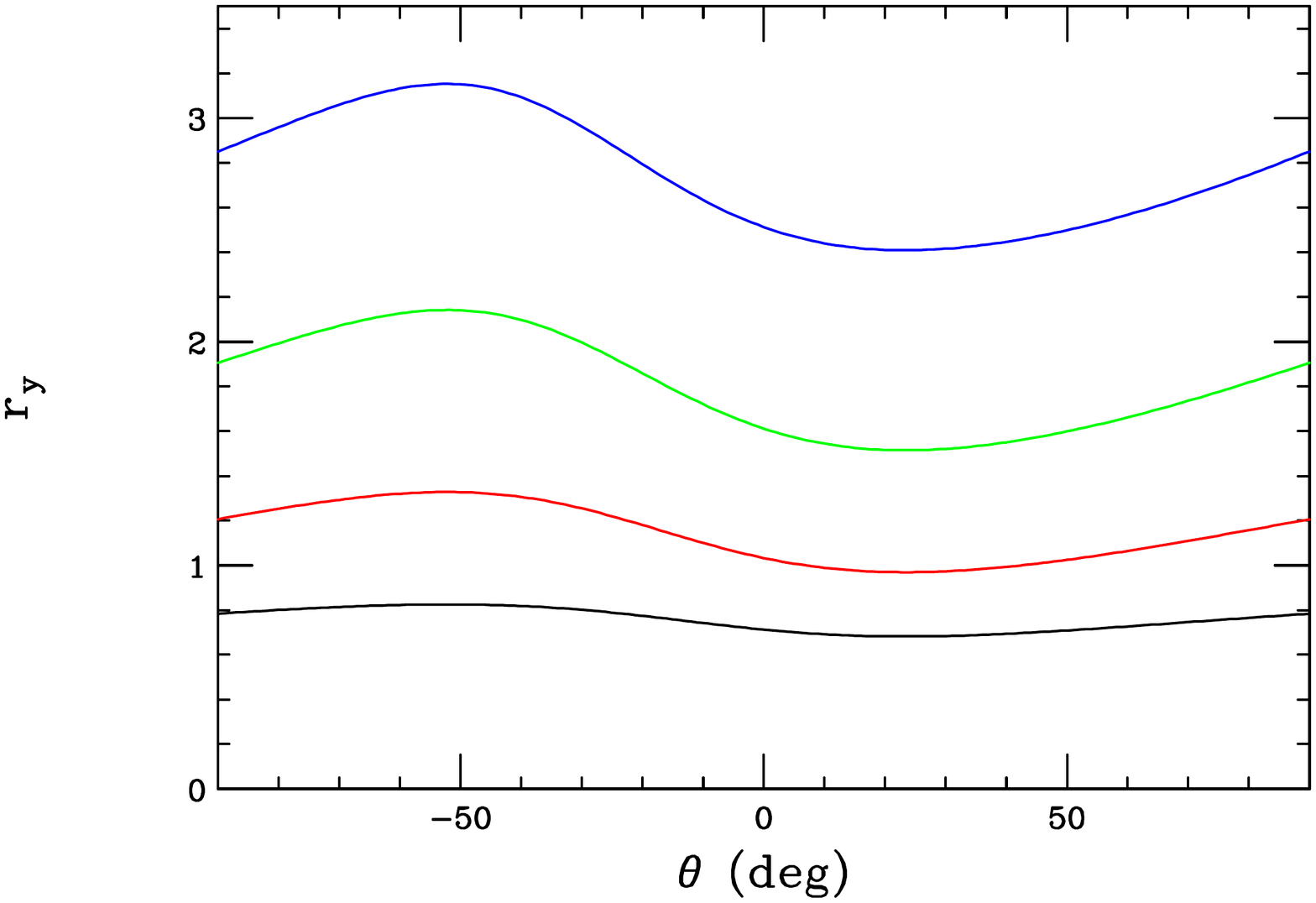}}
\vspace*{-4.0cm}
\centerline{\includegraphics[width=5.0in,angle=90]{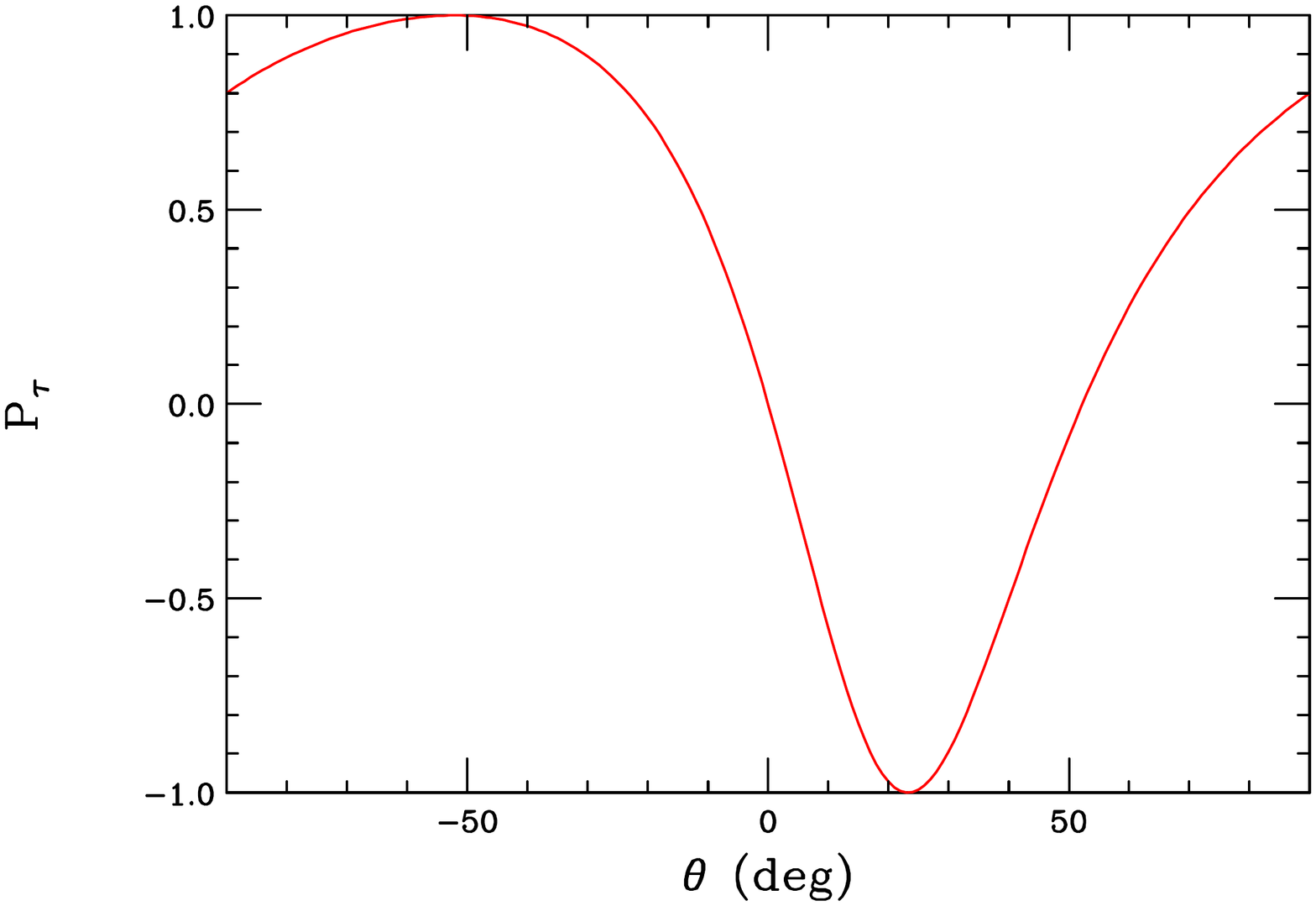}}
\vspace*{-1.50cm}
\caption{(Top) NWA values of $r_y$ in $E_6$ models as above for $Z'$ masses of 5, 10, 15 and 20 TeV (from bottom to top) after the rapidity 
cuts as described in the text have been applied. (Bottom) Tau polarization asymmetry in the NWA for $E_6$ type models.}  
\label{fig08}
\end{figure}

A third possible observable employing the dilepton final state is to construct rapidity ratio{\cite {rev}} in the region near the $Z'$ pole 
which is define via:
\begin{equation}
r_y={{\int^{y_1}_{-y_1}~{{d\sigma} \over {dy}}~dy}\over{\Big[\int^{Y}_{y_1}+\int^{-y_1}_{-Y}~{{d\sigma} \over {dy}}~dy\Big]}}\,.
\end{equation}
This ratio essentially measures the fraction of central rapidity events which depends upon the relative weighting of the $u\bar u$ and 
$d\bar d$ pdfs with the cut $y_1$ defining the central region boundary and $Y=min(2.5,y_{max})$. Note that as the $Z'$ mass increases 
for fixed $y_1$ the ratio $r_y$ will grow significantly due to the kinematic constraints arising from $y_{max}$. For demonstration purposes, 
here we will assume $y_1=0.5$ but this value generally needs to be re-optimized for different values of the $Z'$ mass. To examine the $r_y$ 
sensitivity to $Z'$ coupling variations we present the results shown in Table~\ref{rates3} and Fig.~\ref{fig08} for various models and which 
employ the NWA. For low $Z'$ masses the sensitivity is seen to be rather modest but increases significantly as the mass increases. Of course the 
available statistics to make this measurement decreases significantly for large masses and the statistical error on $r_y$ scales as 
$\sim (1+r_y)(r_y/ N)^{1/2}$ where $N$ is the total number of relevant signal events. For example, this would imply, for a $Z'$ with a mass of 
10 TeV, that $\delta r_y=0.027(0.038,0.046)$ for the SSM($\chi,\eta$) case assuming a luminosity of 5 ab$^{-1}$ whereas for a mass of 15 TeV 
we would instead obtain the values $\delta r_y=0.129(0.189,0.215)$, respectively.  Clearly there is some rather modest model sensitivity 
obtainable using this observable for relatively low $Z'$ masses. We could try getting more information by looking at $r_y$ as a function of 
the dilepton invariant mass, $r_y(M)$, but as Fig.~\ref{fig072} shows the overall behavior of this quantity is much more driven by the the 
shrinking rapidity range (for a fixed $y_1$ cut) as $M$ increases than by the $Z'$ couplings.

\begin{table}
\centering
\begin{tabular}{|l|c|c|c|c|} \hline\hline
Model    & 5  TeV   & 10 TeV  &  15 TeV  & 20 TeV   \\
\hline

SSM        & 0.725  & 1.060 &  1.655 & 2.561  \\
LRM        & 0.740  & 1.096 &  1.713 & 2.625  \\
$\psi$     & 0.712  & 1.033 &  1,612 & 2.513   \\
$\chi$     & 0.783  & 1.206 &  1.905 & 2.851  \\
$\eta$     & 0.691  & 0.987 &  1.542 & 2.438  \\
I          & 0.825  & 1.328 &  2.142 & 3.153  \\
\hline\hline
\end{tabular}
\caption{$r_y$ at $\sqrt s=100$ TeV for different $Z'$ model masses employing the NWA and the rapidity cuts as described in the text.}
\label{rates3}
\end{table}

As a last possibility worth mentioning, one can imagine employing the ditau decay mode of the $Z'$ and then measure the polarization of the $\tau$'s 
themselves\cite {rev}. The lower panel in Fig.~\ref{fig08} shows that this quantity, if measurable, has quite significant coupling sensitivity. 
However, at the 100 TeV FHC, performing this measurement using, \eg, the $\pi$ or $\rho$ decays from the $\tau$ may prove to be rather difficult 
given the rather complex hadronic environment one might naively expect. However this is certainly worthy of further study. 

\begin{figure}[htbp]
\centerline{\includegraphics[width=5.0in,angle=90]{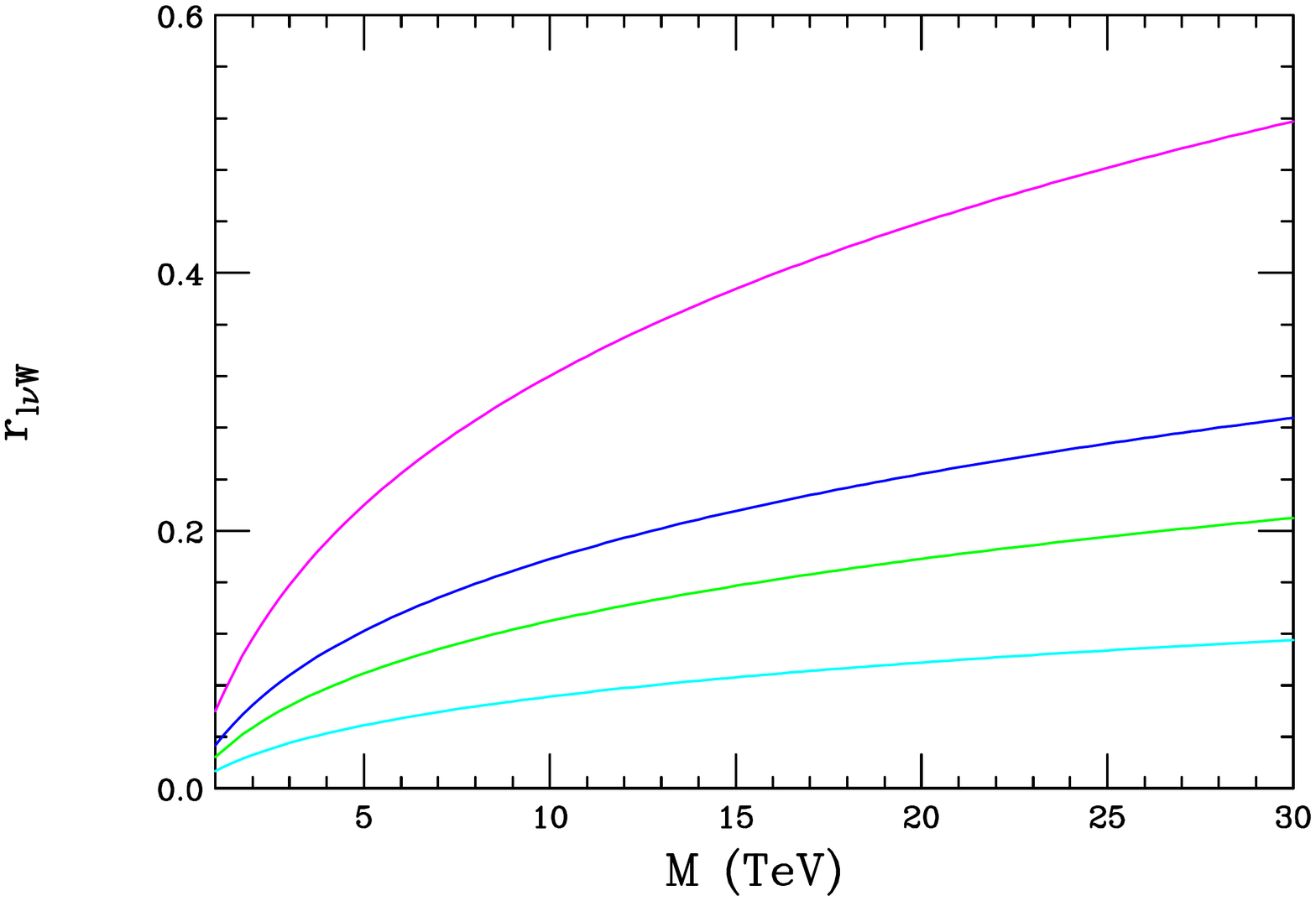}}
\vspace*{-4.0cm}
\centerline{\includegraphics[width=5.0in,angle=90]{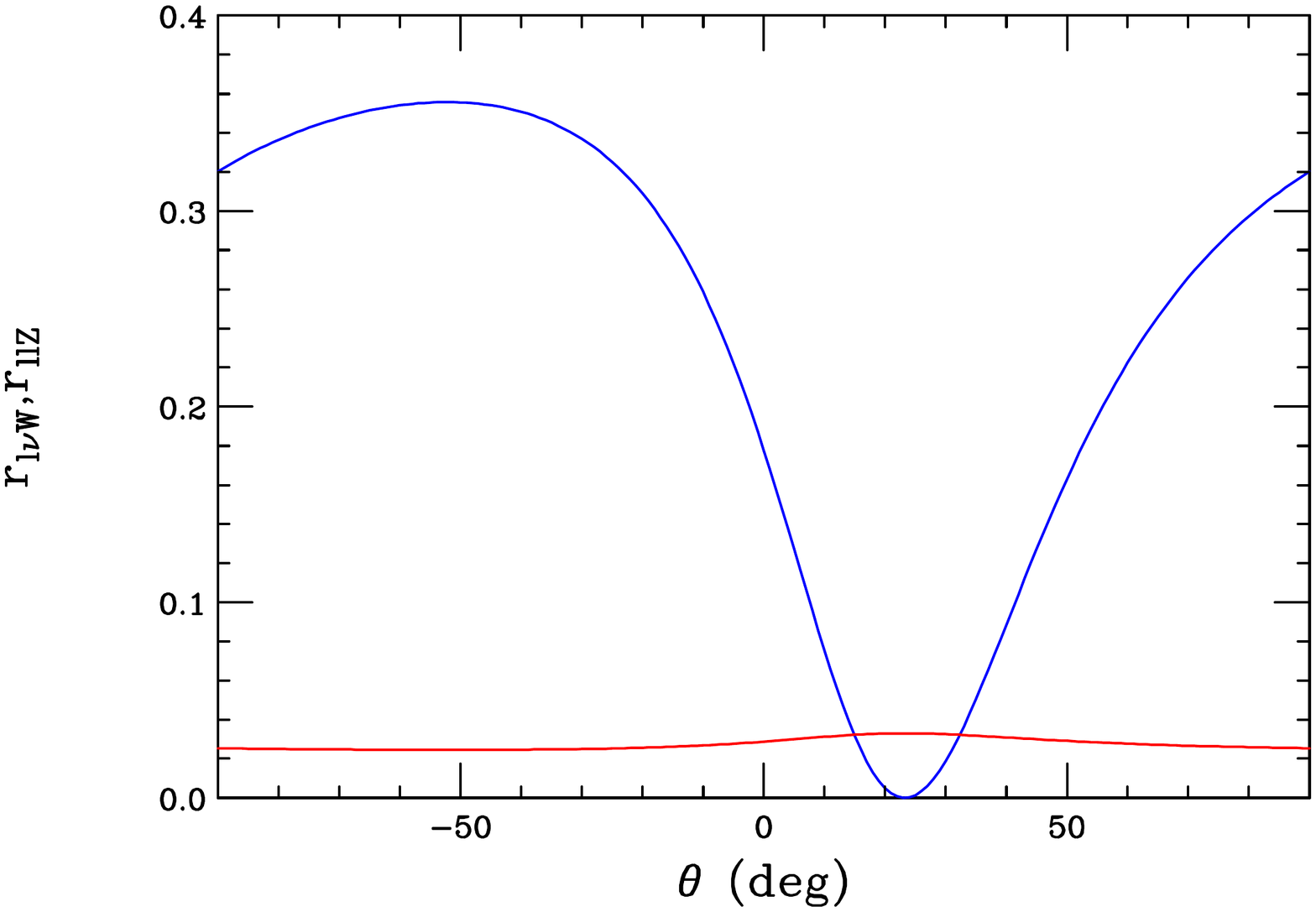}}
\vspace*{-1.50cm}
\caption{(Top) The ratio $r_{l\nu W}$ as a function of the $Z'$ mass for, from top to bottom, the models $\chi$, LRM, $\psi$ and $\eta$. 
(Bottom) The ratios $r_{l\nu W}$ (blue) and $r_{llZ}$ (red) as a function of the $E_6$ parameter $\theta$ assuming a $Z'$ mass of 10 TeV.}
\label{fig09}
\end{figure}

As discussed above, here we have been limiting our discussion to signals employing leptonic triggers so that we will not consider modes such as 
$Z' \to b\bar b$. Thus we are led to consider 3-body decays involving leptons such as $Z'\to l^+l^-Z$ and $Z'\to l^\pm \nu W^\mp$ and, 
in particular the ratios of the partial widths for these reactions relative to that for $Z\to l^+l^-$ denoted as $r_{llZ}$ and $r_{l\nu W}$ 
in the literature\cite {rev}. As discussed above, we employ these ratios to remove the potential sensitivity to non-SM $Z'$ decays. 
Assuming that the $Z'W^+W^-$ coupling is absent at tree-level (which it is in the GUT-type models under 
consideration in the absence of $Z-Z'$ mixing which we assume here), these processes occur by the emission of $W/Z$ gauge bosons from one 
of the lepton legs in ordinary $Z'$ dilepton decays.  At the LHC, for $Z'$ masses $\sim 1$ TeV, these ratios were shown to be useful for 
coupling determinations but suffered from small statistics. We note in Fig.~\ref{fig09}, however, that $r_{l\nu W}$ (and similarly $r_{llZ}$) 
grow like $\sim \rm log^2 (M_{Z'}/M_W)$ due to the near infrared and co-linear sigularities in the relevant graphs. This renders them potentially 
far more useful than at lower collision energies although clearly $r_{l\nu W}$ should be the focus of our attention. For example, for a $Z'$ 
mass of 15 TeV in the LRM the cross section for the $l^\pm \nu W^\mp$ final state is $\sim 50$ ab. One difficulty at these energies is the 
rather large boost of the final state $W,Z$ and its small opening angle with respect to the lepton from which it was emitted, \ie, isolation 
issues. Further, if the $W,Z$ are found through their dijet decay modes these two jets will not allow for $W,Z$ separation and will also 
likely appear as a single jet. This final state warrants further study at the FHC.

\begin{figure}[htbp]
\centerline{\includegraphics[width=5.0in,angle=90]{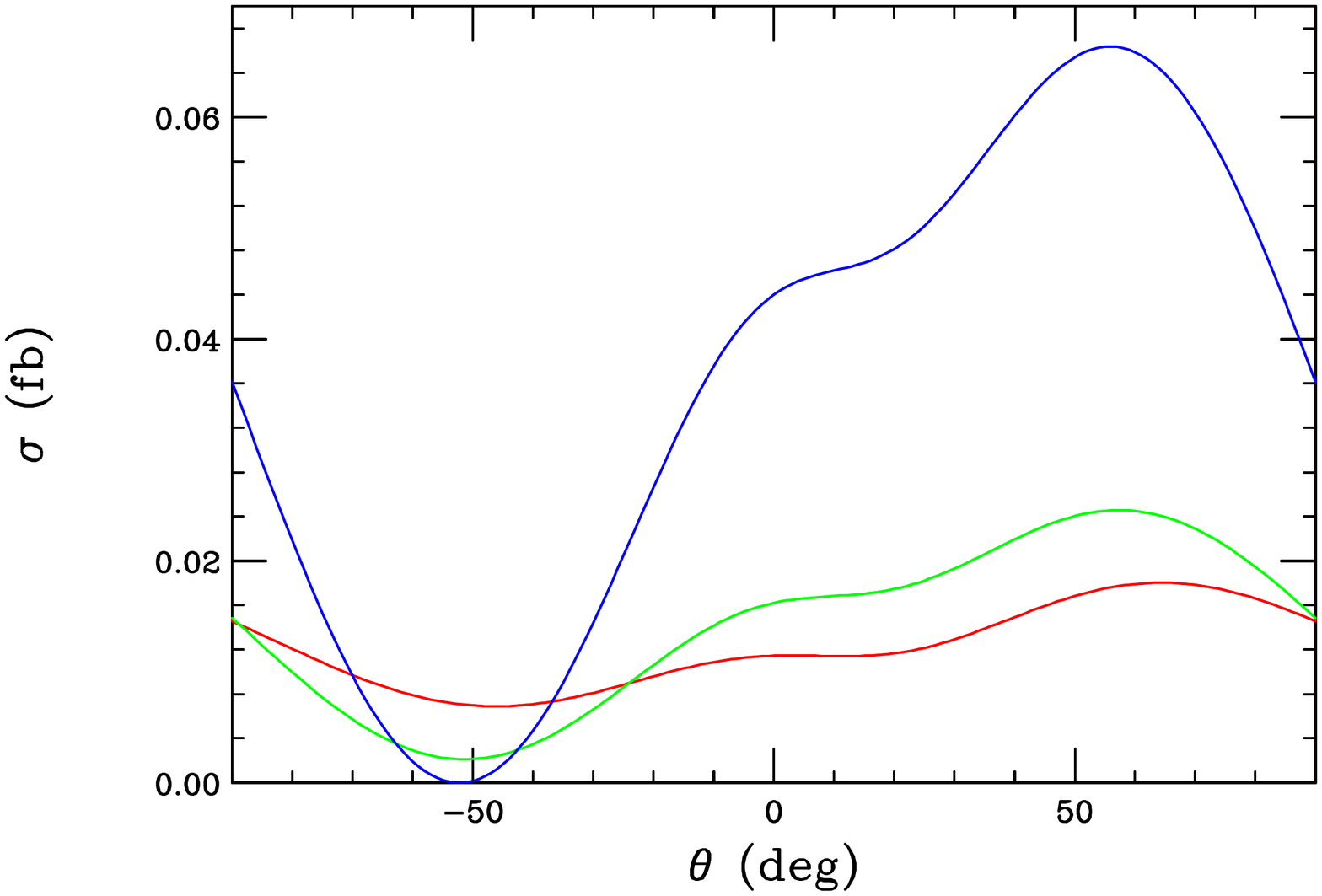}}
\vspace*{-4.0cm}
\centerline{\includegraphics[width=5.0in,angle=90]{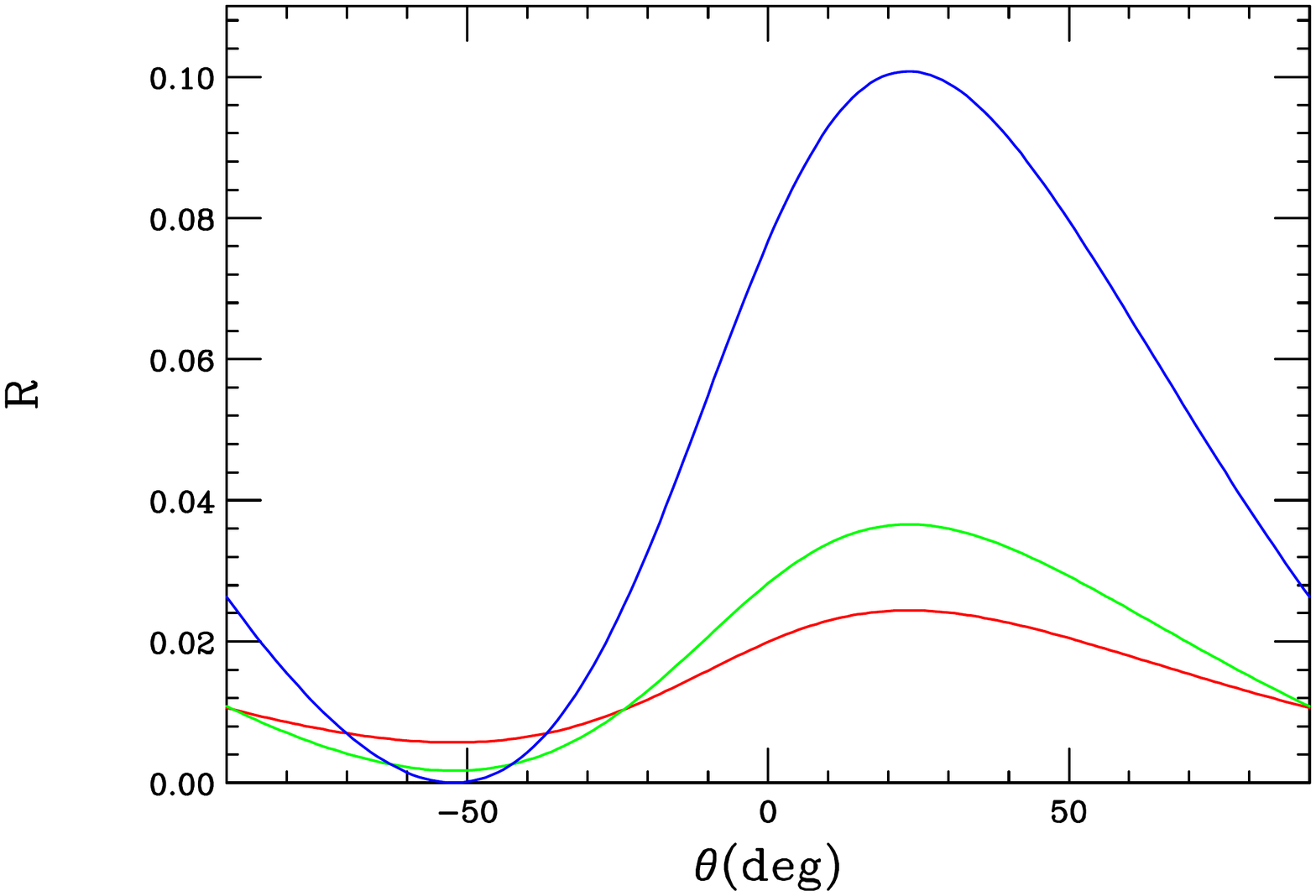}}
\vspace*{-1.50cm}
\caption{(Top) Associated $VZ'$ production cross section followed by $Z' \to l^+l^-$ for $V=W^\pm$(blue), $V=Z$(green) and $V=\gamma$ with 
$p_T>100$ GeV (red) in the $E_6$ scenario taking a $Z'$ mass of 10 TeV. (Bottom) The corresponding ratios $R_V$.}
\label{fig10}
\end{figure}

As a final potential probe of $Z'$ couplings, we can consider the cross sections for the associated production of the $Z'$, observed via its 
leptonic decay mode as usual, together with some other SM gauge field, a high-$p_T ~\gamma$, $W$ or $Z${\cite {rev}} produced via `initial 
state radiation'. 
Scaling these cross sections to that for ordinary $Z'\to l^+l^-$ production yields a set of observables, $R_{V(=\gamma,W,Z)}$, that are 
immune to the possible existence of non-SM decay modes. Fig.~\ref{fig10} shows both the production cross sections for these final states 
as well as the associated ratios $R_V$ assuming a 10 TeV $Z'$ mass at the 100 TeV FHC. Note that in the $V=\gamma$ case, a $p_T$ cut of 100 
GeV has been applied to the final state photon. (The above comments about the $W$ and $Z$ final state identification in dijets will also apply 
here as well.) These results show that 
the associated production observables will suffer from somewhat low statistics if only$\sim 1$ ab$^{-1}$ of integrated luminosity is available 
at the FHC but they still warrant further study at the FHC.

\section{New $W'$ Bosons}

We now turn our attention to new $W'$ gauge bosons which are generally far less well studied than are the $Z'$.  
As in the $Z'$ case the first issue to address is the $W'$ production cross section at the 100 TeV FHC; as for the $Z'$ we limit ourselves to 
final states involving lepton tags so that the primary discovery channel is $W'\to l\nu$ with the $\nu$ appearing as a large amount of MET. 
All calculations for the $W'$ will be performed in the same manner as in the $Z'$ case above making use of the same inputs and assumptions. 

\begin{figure}[htbp]
\centerline{\includegraphics[width=5.0in,angle=90]{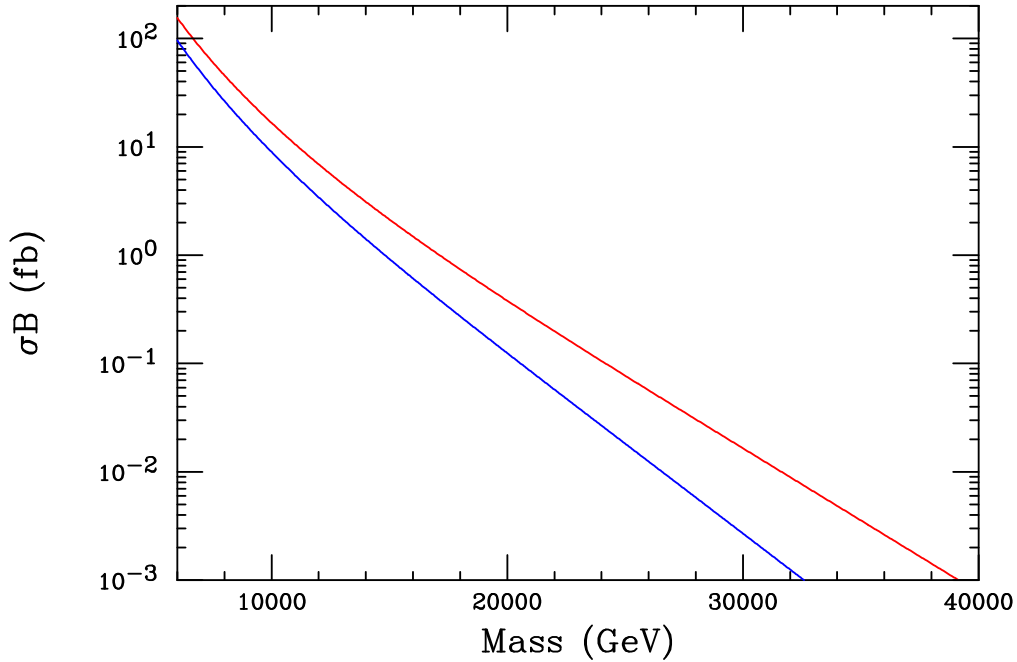}}
\vspace*{-0.90cm}
\caption{$W'^\pm \to l^\pm\nu$ production cross section in the NWA as a function of the $W'$ mass at both $\sqrt s=80,100$ TeV.}
\label{fig11}
\end{figure}

Here we face an immediate issue: in essentially every model in the literature the couplings of the $W'$ to SM fermions are either purely LH 
or RH. {\it Assuming} that neutrinos are either Dirac fermions or are light and detector-stable Majorana fermions, so that this same 
final state occurs in both cases and that, apart from phases, the 
RH and LH CKM mixing matrices are the same (which is a likely occurrence in, \eg, the LRM) then in the NWA limit the $W'$ production cross 
section is independent of this helicity choice\cite{Rizzo:2007xs}. The resulting cross section is shown in Fig.~\ref{fig11} for both 
$\sqrt s=80$ and 100 TeV for comparison purposes. Here we can observe several things (that will be seen in more detail below): ($i$) At 
$\sqrt s=80(100)$ GeV the discovery reach is $\simeq 20.6(31.6)$ TeV while ($ii$) the $95\%$ CL exclusion limit is found to be 26.6(35.5) 
TeV. If the integrated luminosity is increased by a factor of 3(10) these numbers all increase by roughly 3.6(7.5) TeV. 
($iii$) At 100 TeV, if exotic modes are present to reduce the $W'$ leptonic branching fraction by a factor of 2, then the corresponding 
discovery reach is reduced by $\simeq 2.2$ TeV. It is important to remember that since the $W'$ couplings are either LH or RH, determining 
{\it which} of these is realized is the primary goal after the $W'$ discovery{\footnote {As discussed above we only consider leptonic modes 
here;see however\cite{Gopalakrishna:2010xm}, \cite{Han:2012vk} and \cite{Berger:2011hn} for other possibilities}} .

\begin{figure}[htbp]
\centerline{\includegraphics[width=3.5in,angle=90]{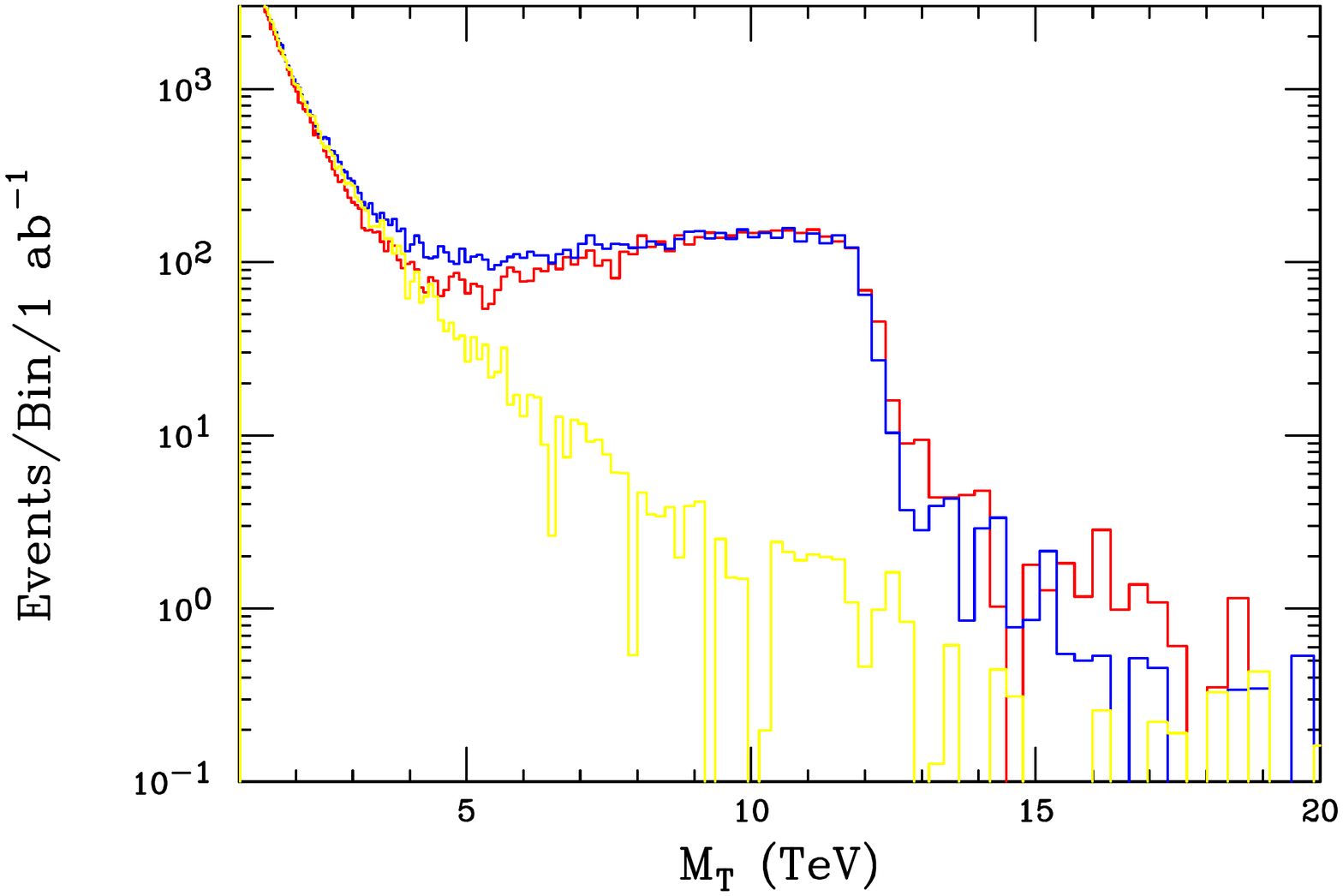}
\hspace {-1.2cm}
\includegraphics[width=3.5in,angle=90]{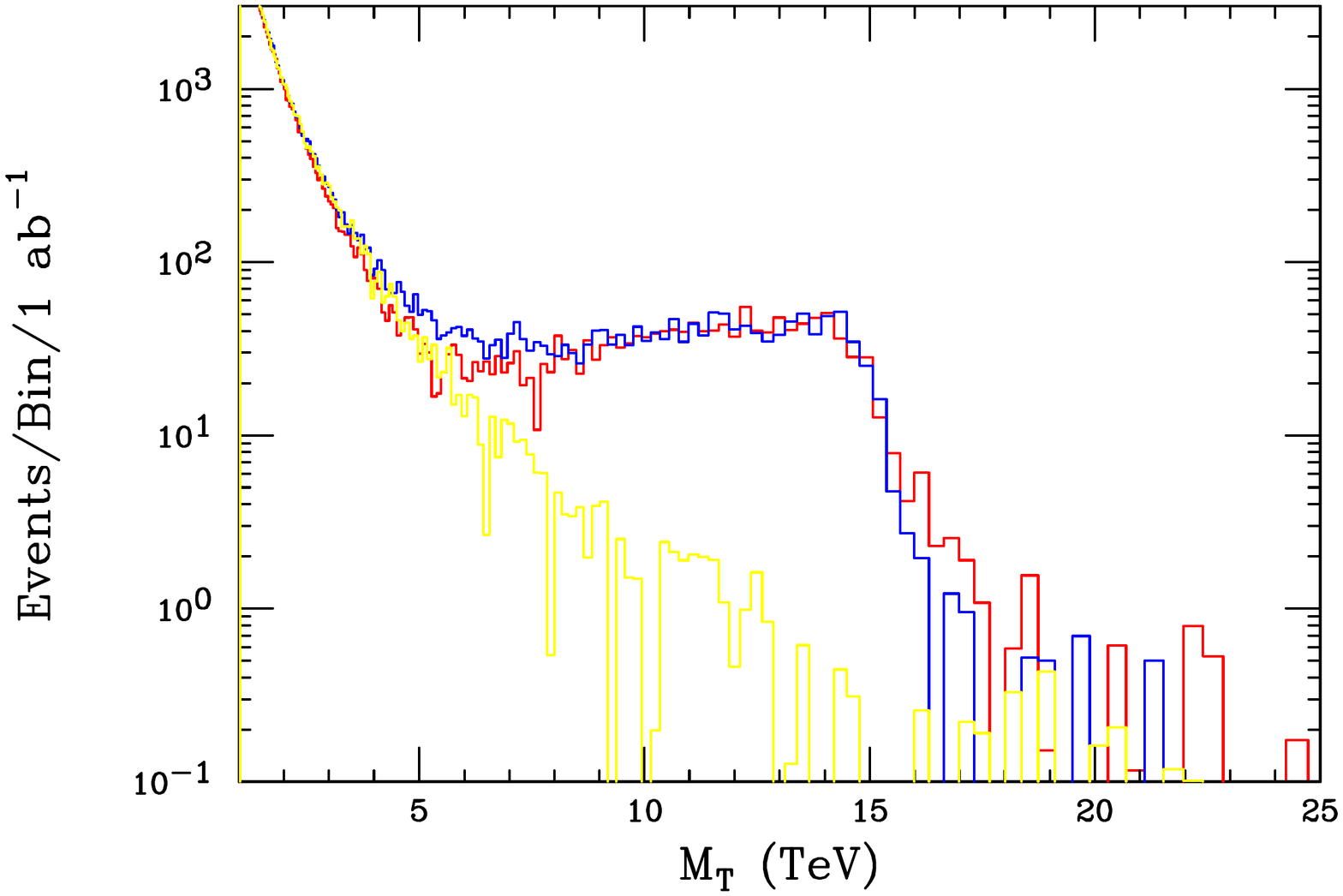}}
\vspace*{-1.0cm}
\centerline{\includegraphics[width=3.5in,angle=90]{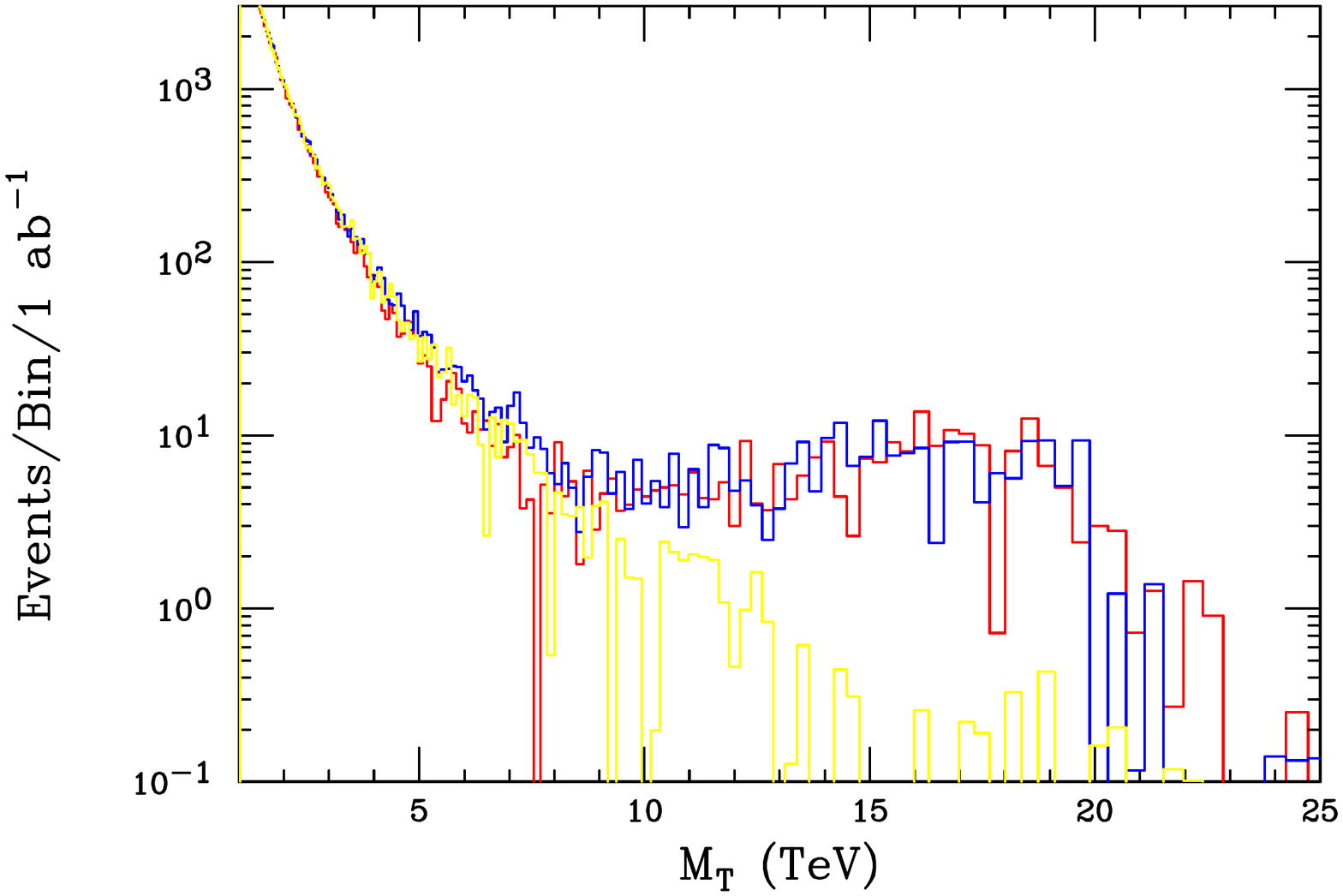}
\hspace {-1.2cm}
\includegraphics[width=3.5in,angle=90]{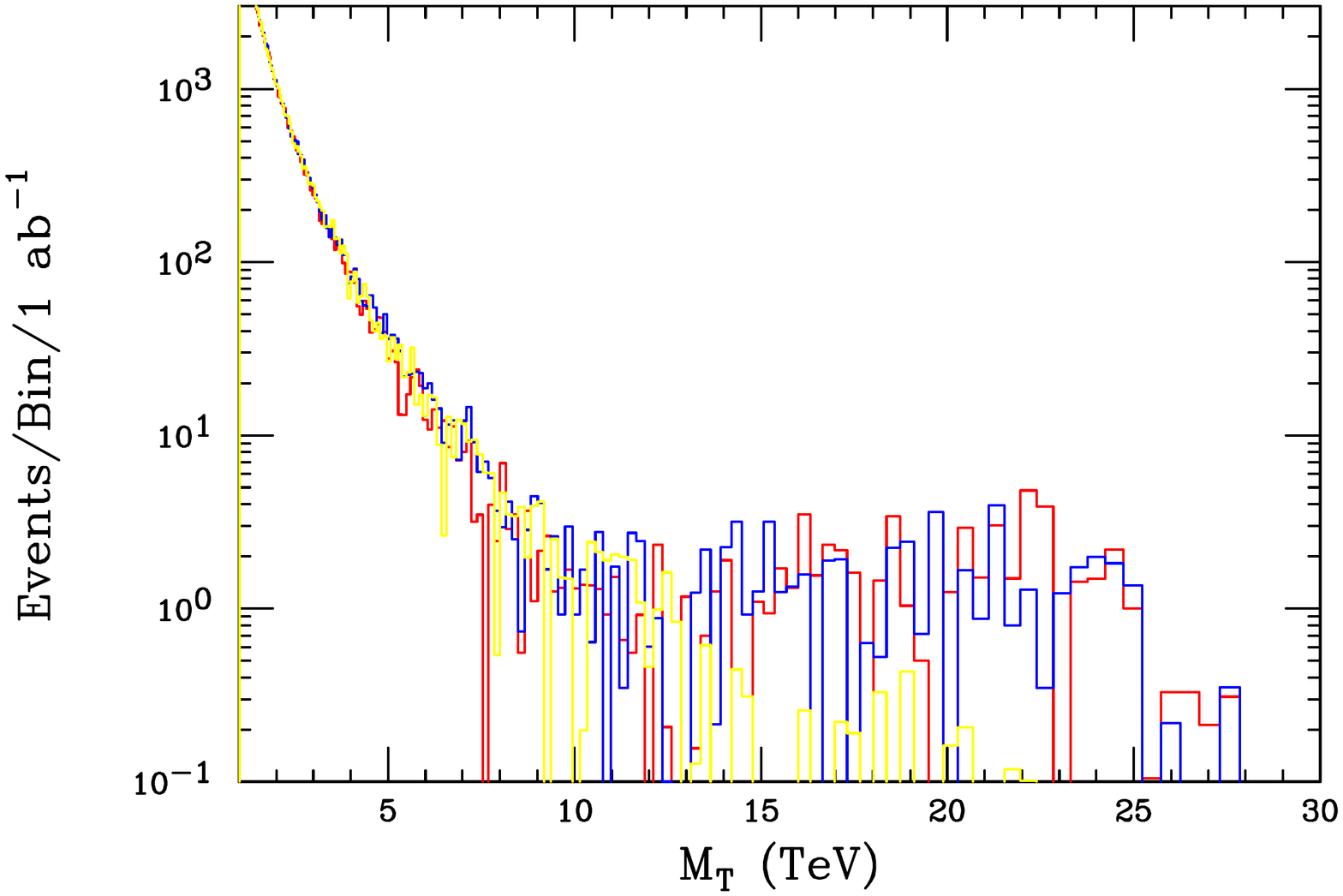}}
\vspace*{-1.00cm}
\caption{Transverse mass distributions for $W'$ production at $\sqrt s=100$ TeV as discussed in the text. The SM background is shown in yellow 
whereas the signal for a LH(RH) coupled $W'$ is shown in red(blue). The upper left(right) panel is for a $W'$ mass of 12(15) TeV whereas the 
lower left(right) panel corresponds to a mass of 20(25) TeV. An integrated luminosity of 1 ab$^{-1}$ has been assumed in obtaining these results.}
\label{fig12}
\end{figure}

\begin{figure}[htbp]
\centerline{\includegraphics[width=3.5in,angle=90]{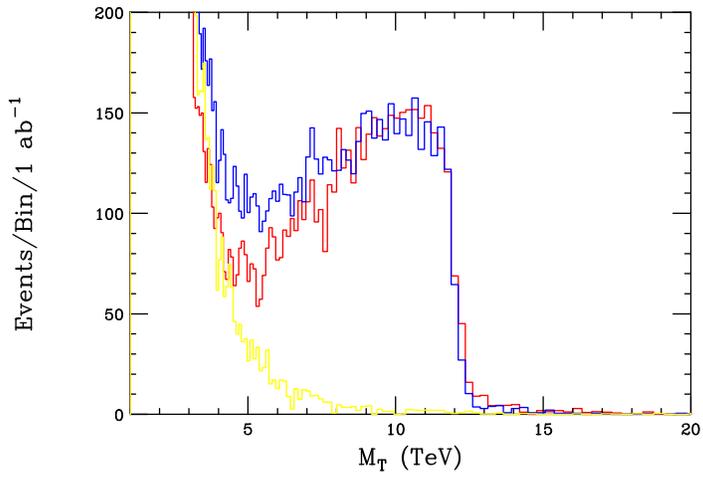}
\hspace {-1.2cm}
\includegraphics[width=3.5in,angle=90]{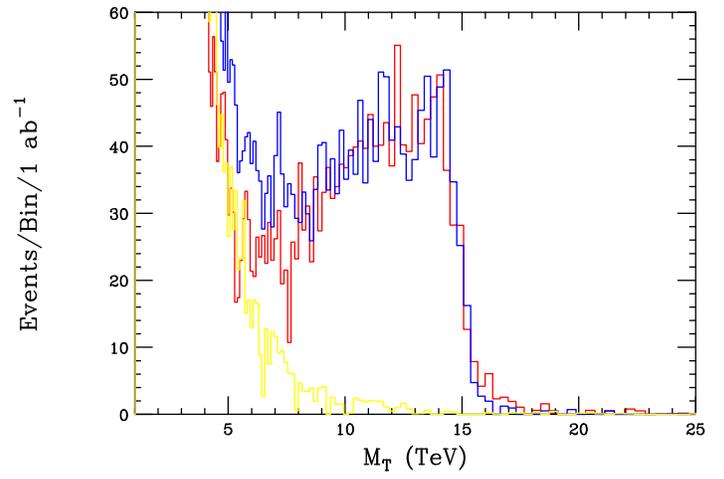}}
\vspace*{-1.0cm}
\centerline{\includegraphics[width=3.5in,angle=90]{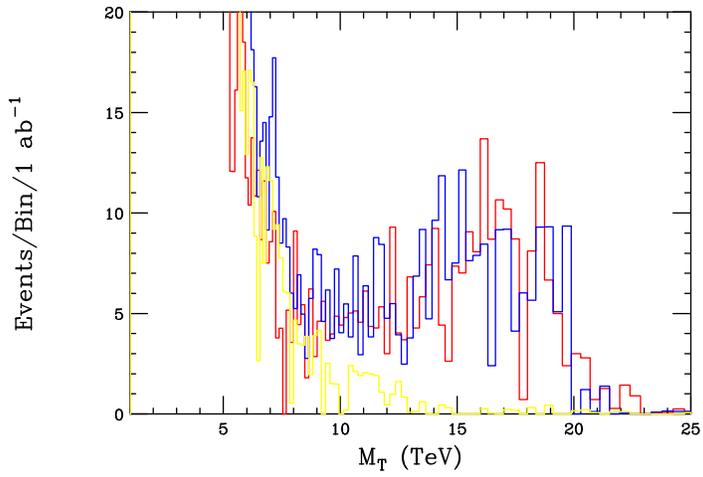}
\hspace {-1.2cm}
\includegraphics[width=3.5in,angle=90]{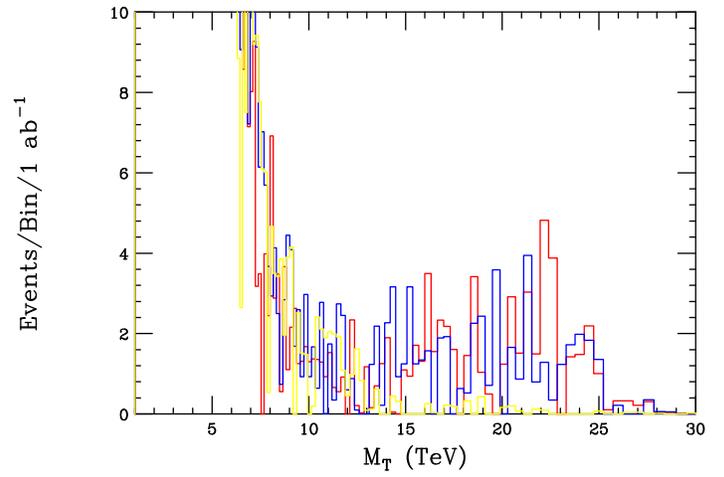}}
\vspace*{-1.00cm}
\caption{Same as in the previous Figure but now showing with a linear vertical scale.}
\label{fig13}
\end{figure}

Of course the actual way that a $W'$ would be discovered would be via the observation of an excess in the number of events at the high end of 
the transverse mass ($M_T$) distribution formed from the charged lepton and the MET as in the case of the SM $W$. (Here we will need to assume that 
the FHC detector(s) can easily identify large amounts of MET with ATLAS-like resolution.) To this end it is instructive 
to examine these distributions for different values of the $W'$ mass assuming either LH or RH couplings to the usual SM fermions. The results 
for $W'$ masses of 12, 15, 20 and 25 TeV are shown in Figs.~\ref{fig12} and ~\ref{fig13} on log and linear scales respectively{\footnote {Here we 
display both log and linear scale event rate plots as we can learn something from both of these projections.}}.  Several things are 
immediately apparent: 
($i$) the $W'$ for all these masses, whether LH or RH, is clearly visible above the SM background in a statistically significant way. 
($ii$) The behavior of the $M_T$ distribution in the neighborhood of the Jacobian peak is the same (within statistics) whether the $W'$ is LH 
or RH. This conforms with our expectations from the NWA analysis above. ($iii$) The overall shapes of the $M_T$ distributions for the LH and 
RH $W'$ couplings are {\it not} the same\cite{Rizzo:2007xs} and CMS\cite{CMS:2013rca} in particular has exploited this fact as part of their $W'$ 
search. This difference is particularly noticeable in two regions: for $M_T \sim 0.4M_{W'}$ and above 
the Jacobian peak. In the lower mass region we see that the LH result lies below the RH one whereas above the Jacobian peak the reverse is true. 
This difference is due to the interference (or lack thereof) between the $W'$ and the SM $W$ amplitudes. In the LH case, these amplitudes have 
opposite signs below the Jacobian peak but add constructively above it. In the RH $W'$ case, however, there is no interference with the SM $W$ 
amplitude since the SM fermions in both the initial and final states are treated as massless. 
It is clear from these Figures that this difference in the $M_T$ distributions is apparently visible (at this assumed integrated luminosity) for 
$W'$ masses possibly as large as $\sim 20$ TeV and may be potentially so even for somewhat larger masses even though all these values are still 
significantly below the corresponding discovery reach. Clearly no difference would be observable for a mass of 30 TeV although an excess of events 
would still be seen in both helicity cases. To get a clearer idea of the size of this cross section difference due to the $W'$ coupling helicity  
(since the Figures can be sometimes misleading due to the small $M_T$ bin size) let us first imagine that a $W'$ has been seen at a given mass 
through its Jacobian peak. One can then just count the total number of events in the $M_T$ distribution in the region where the interference is 
most important, $M_T \geq 0.25 M_{W'}$, say, and compare the expectations in the two cases. For a LH(RH) $W'$ with a mass of 12 TeV and an 
integrated luminosity of 1 ab$^{-1}$ one obtains 8153(9818) events which certainly differ by more than $5\sigma$(statistical errors only). 
Similarly, for $W'$ masses of 15, 20 and 25 TeV, one correspondingly obtains 2669(3319), 592(771) and 175(208) events, respectfully, indicating 
good separation up to masses of $\sim 20$ TeV for these two coupling helicities. Of course, increased integrated luminosity will clarify only these 
issues further.

Although the shape of the $M_T$ distribution is a powerful probe of the handedness ($h_{W'}=1(-1)$ for LH(RH) couplings) of the $W'$, it would 
very useful to have other observables available to also determine this quantity. In the $Z'$ discussion above we saw that $A_{FB}$ was a useful 
tool to get at the corresponding couplings. Here, since the SM fermions couple to the $W'$ as $\bar f \gamma_\mu (1-h_{W'}\gamma_5)f'W^\mu$, 
the corresponding asymmetry on the $W'$ pole in the NWA would behave as 
\begin{equation}
A_{FB} \sim {{h_{W'}^2}\over {(1+h_{W'}^2)^2}}\,,
\end{equation}
which trivially gives the same result for both helicities and is thus useless for present purposes. Again, as in the $Z'$ case, one might attempt 
to measure the polarization of the $\tau$ in $W' \to \tau \nu_\tau$ but it is likely to be even more difficult in this case due to the additional MET 
and busy detector environment. Building further on our $Z'$ experience we could try to examine both the associated $W'W$ production process (which 
would likely have $Z$ contamination if $W\to jj$ is used) or the rare bremsstrahlung-like decay $W' \to l^+l^- W$ (here cutting away the $Z$ pole 
region) for helicity sensitivity. Naively, if the $W'$ were purely RH both these processes would be absent at tree-level so that just their 
simple observation would inform us that the $W'$ is LH. However, since a traditional $W'$ only arises from an additional $SU(2)$ group factor, it 
is always accompanied by a $Z'$ due to gauge invariance so that vertices such as $W'Z'W$ and/or $W'ZW$ might be present which will not necessarily 
turn on or off depending upon the helicity of the $W'$ coupling to the SM fermions{\footnote {However, as discussed in detail in 
Ref.~\cite{Rizzo:2007xs}, there are no known models with a RH $W'$ where these potentially dangerous vertices are not suppressed by gauge boson 
mixing angles.}}. Thus some care would be required in interpreting the observation of either of these processes and they both warrant further 
investigation.

\begin{figure}[htbp]
\centerline{\includegraphics[width=5.0in,angle=90]{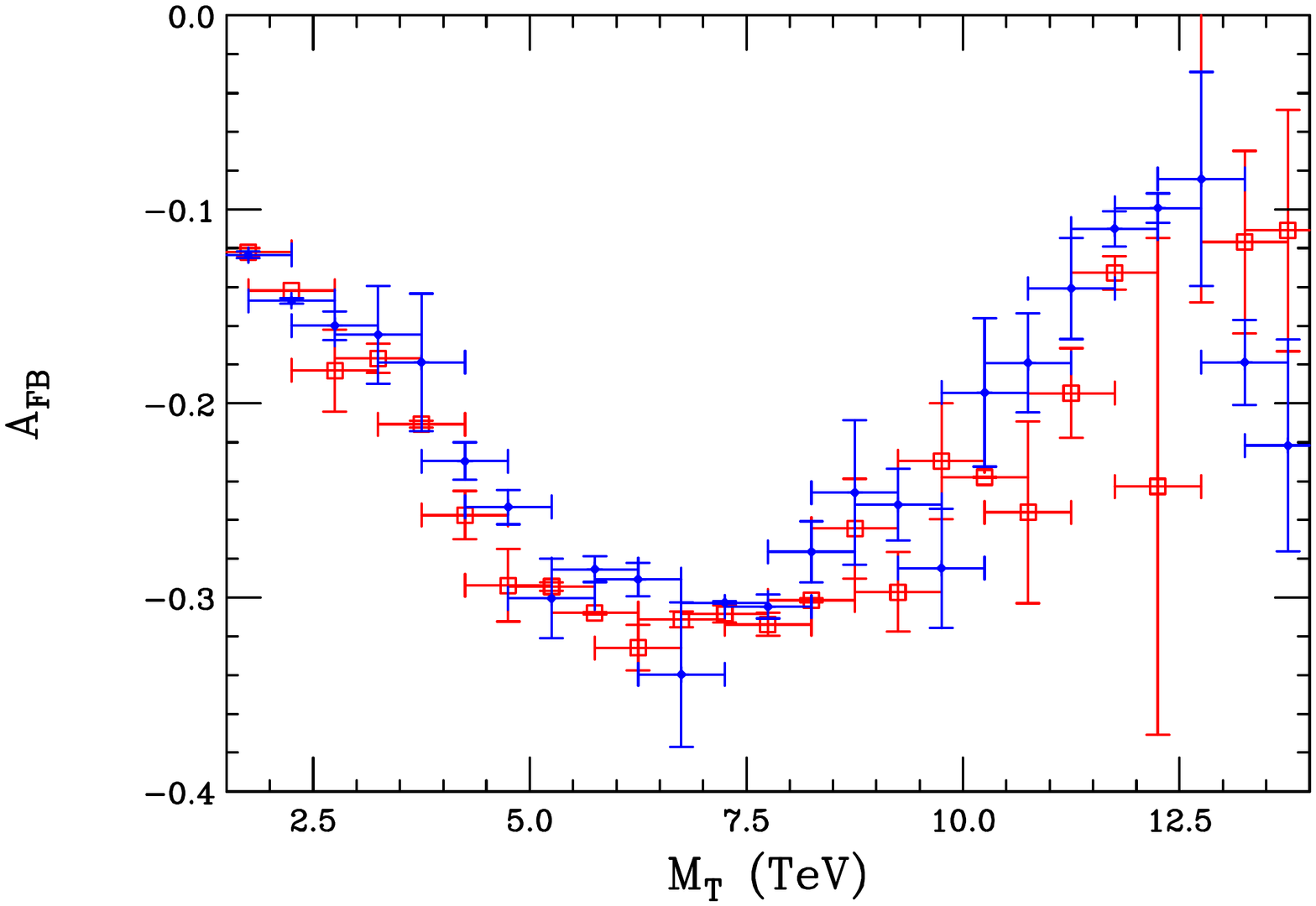}}
\vspace*{-4.0cm}
\centerline{\includegraphics[width=5.0in,angle=90]{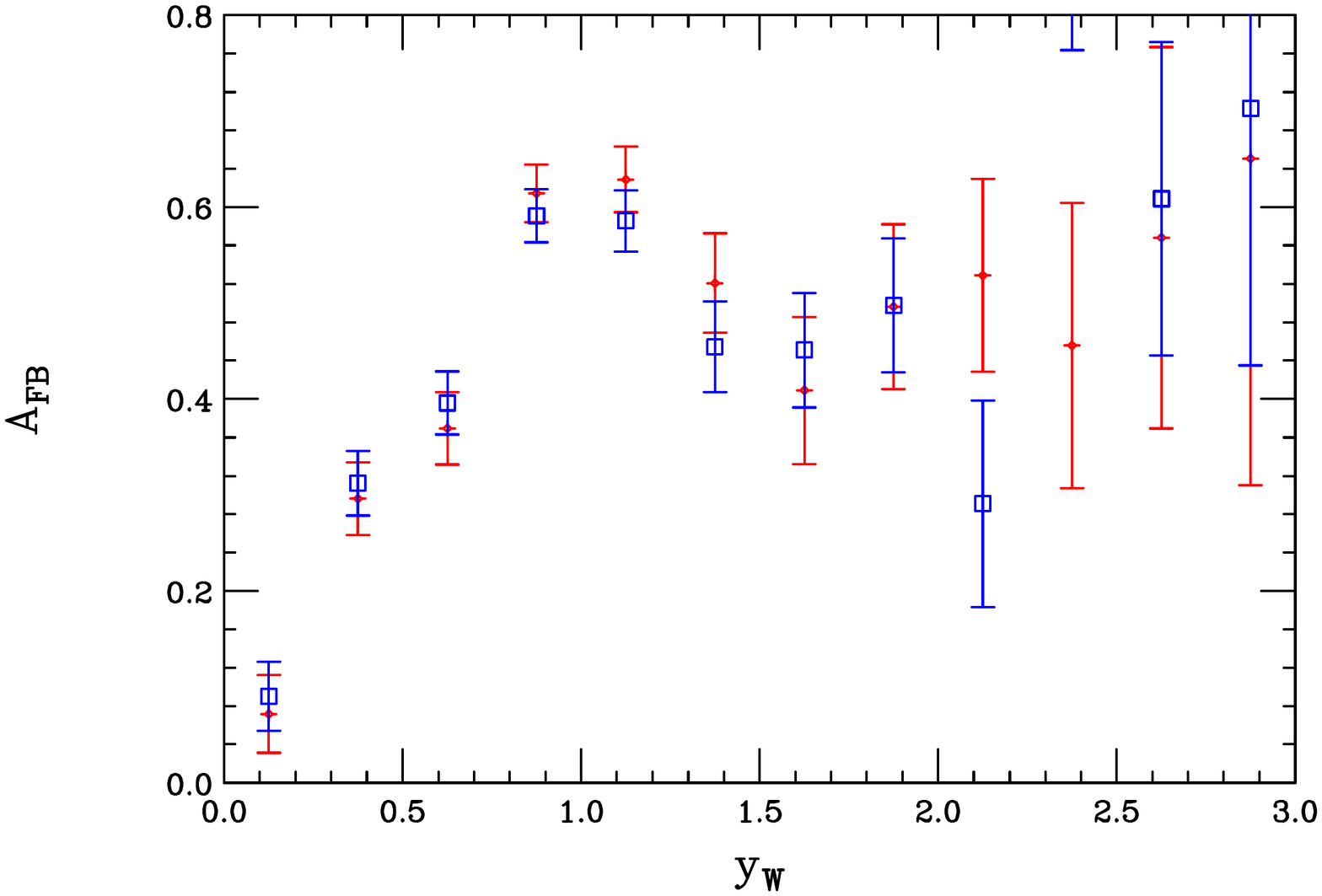}}
\vspace*{-1.50cm}
\caption{(Top) Rapidity integrated values of $A_{FB}$ for the sum of $W'^+$ and $W'^-$ production as a function of the transverse mass 
assuming a $W'$ mass of 12 TeV and a luminosity of 5 ab$^{-1}$ integrated over 500 GeV $M_T$ bins. Here the LH(RH) helicity result is 
shown in red(blue). (Bottom) 
$A_{FB}(y_{W})$ for $y_W \geq 0$ and integrated over the range $3\leq M_T \leq 9$ TeV for a $W'$ mass of 12 TeV and a luminosity of 5 ab$^{-1}$. 
The red(blue) points are for the LH(RH) $W'$ coupling.}
\label{fig14}
\end{figure}

Since these NWA-like processes do not appear to be 
very promising, we turn to a number of asymmetry distributions that one can form using the $l+$MET 
discovery mode of the $W'$. Since we already know about the possible $W-W'$ interference as a probe of the coupling helicities it is worthwhile 
to take advantage of this. Although we've shown that $A_{FB}$ near the Jacobian peak region is insensitive to this helicity choice we might 
wonder if we can learn something by examining lower $M_T$ values where this interference is certainly most relevant. The asymmetry predictions 
for $W'^+$ and for $W'^-$ are found to be slightly different (due to the pdfs) but here we combine them, weighted by their associated cross 
sections,  to increase statistical sensitivity while still integrating over the rapidity. As in the $Z'$ 
case we impose a minimum rapidity cut here in an attempt to define the initial quark direction so that the angle $\theta^*$ can be determined. 
We assume, as we did in the $Z'$ case, that a mis-identification of this angle can be corrected for statistically using Monte Carlo. The top 
panel of Fig.~\ref{fig14} show the results of this calculation for $A_{FB}(M_T)$ employing 5 ab$^{-1}$ and a relatively light $W'$ of mass 12 TeV. 
Clearly there seems to be very little power here to resolve the LH vs. RH dichotomy with this level of statistics. Alternatively we can focus on the 
expected interference region, here taking $M_T$ in the 3-9 TeV range and examine the rapidity dependence of the $A_{FB}$ distribution of the $W'$ 
itself. Since this distribution is odd in $y_W$ we fold it over to increase statistics. We also can again combine the results for $W'^+$ and $W'^-$ 
by a sign flip and event weighting to increase the statistics even further. Unfortunately, as the lower panel of Fig.~\ref{fig14} shows, this 
observable assuming even this large value of the integrated luminosity cannot distinguish between the two helicity possibilities even for a 
relatively light $W'$ of mass 12 TeV.

\begin{figure}[htbp]
\centerline{\includegraphics[width=5.0in,angle=90]{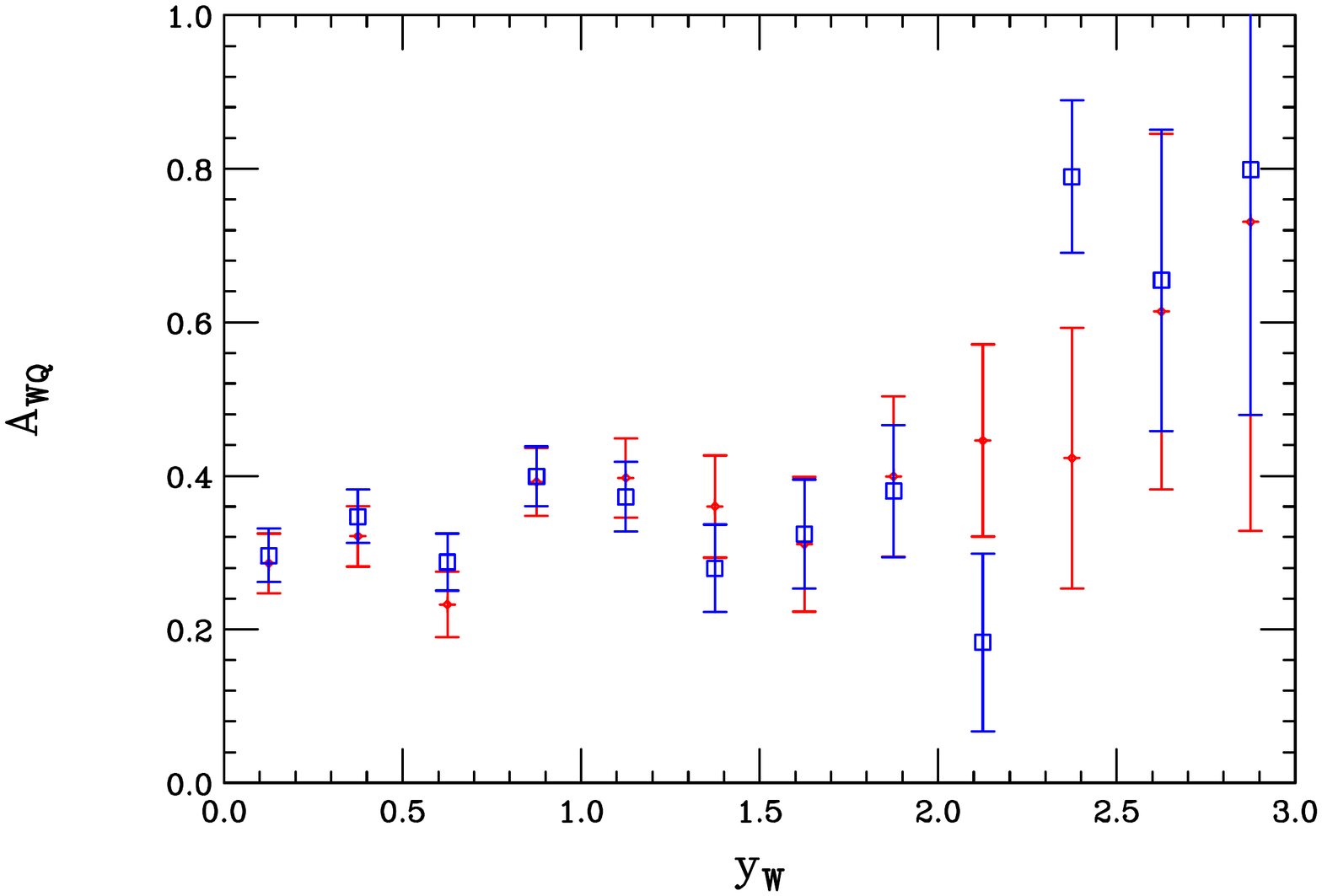}}
\vspace*{-4.0cm}
\centerline{\includegraphics[width=5.0in,angle=90]{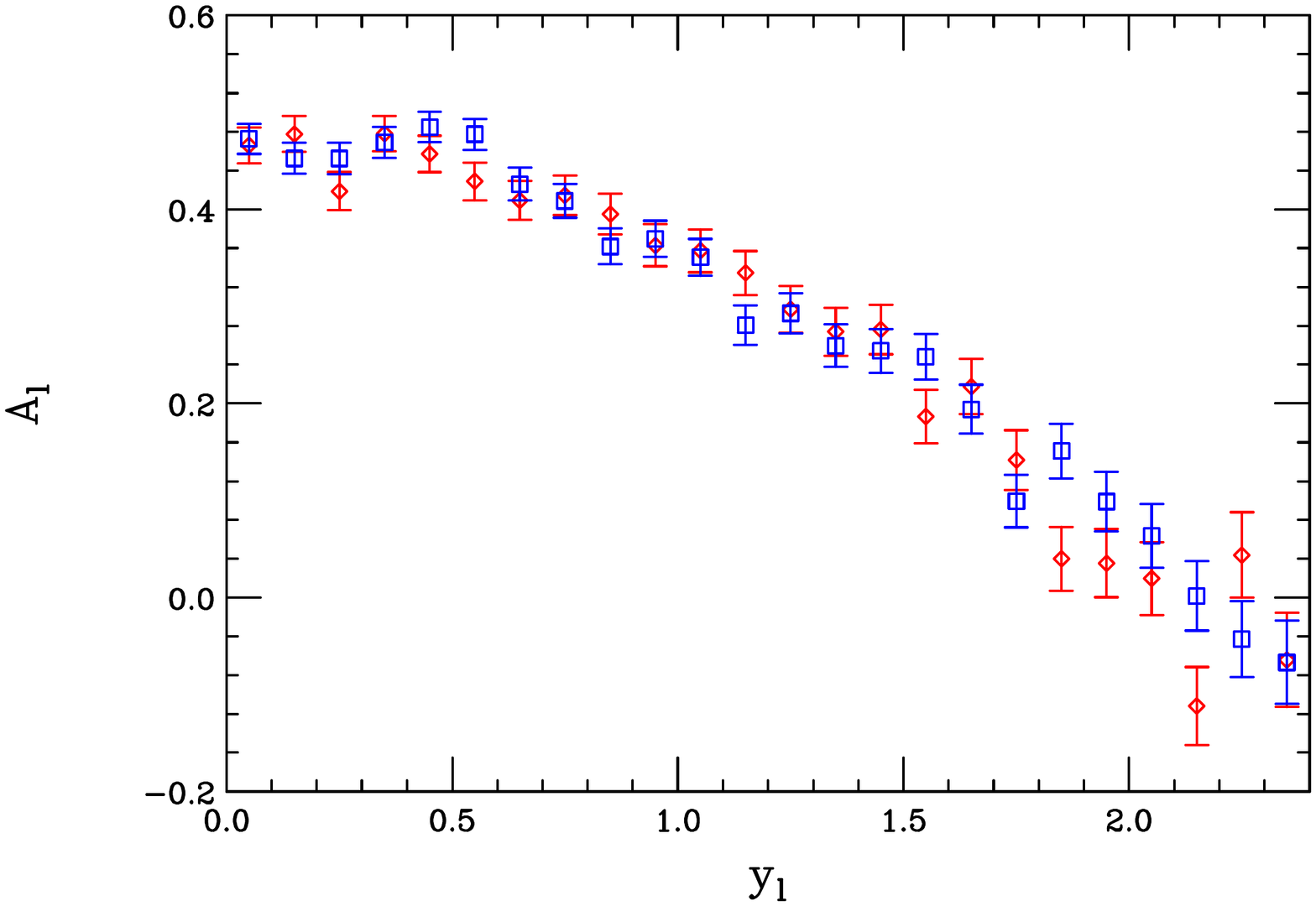}}
\vspace*{-1.50cm}
\caption{(Top) The charge asymmetry as a function of the $W'$ rapidity, $A_{WQ}(y_W)$, integrated over the range $3\leq M_T \leq 9$ TeV for 
a $W'$ mass of 12 TeV and a luminosity of 5 ab$^{-1}$. The red(blue) points are for the LH(RH) $W'$ coupling. (Bottom) The lepton asymmetry, 
$A_\ell(y_\ell)$ as a function of the lepton rapidity for the same physical situation and color labeling.}
\label{fig15}
\end{figure}

There still remains other asymmetries which we can form with this same final state: we first consider is the charge asymmetry, $A_{WQ}(y_W)$:
\begin{equation}
A_{WQ}(y_W)={{N_+(y_W)-N_-(y_W)}\over {N_+(y_W)+N_-(y_W)}}\,,
\end{equation}
where $N_\pm(y_W)$ are the number of events with $W'$s of a specific charge $\pm$ in given rapidity bin. Note that at pp colliders, 
$A_{WQ}(y_W)$ is symmetric under $y_W \to -y_W$ so that we can again fold the distribution around $y_W=0$ to increase statistics. The upper 
panel in Fig.~\ref{fig15} shows this distribution, integrated over the interference region $3 \leq M_T \leq 9$ TeV, assuming $M_{W'}=12$ TeV 
and assuming a luminosity of 5 $ab^{-1}$. It is clear that at this level of statistics the two distributions are reasonably indistinguishable. 
As a final asymmetry possibility, we consider is the rapidity asymmetry for the final state charged leptons themselves, $A_\ell(y_\ell)$:
\begin{equation}
A_{\ell}(y_\ell)={{N_+(y_\ell)-N_-(y_\ell)}\over {N_+(y_\ell)+N_-(y_\ell)}}\,,
\end{equation}
which is an even function of $y_\ell$ so that this asymmetry too can be folded around $y_\ell=0$. The resulting distribution can be seen 
in the lower panel of Fig.~\ref{fig15} for the same $W'$ mass and integrated luminosity. Here we again see hardly any model differentiation 
between the two helicity choices. 

From this general discussion of possibly asymmetries that one can form employing this leptonic final state we can conclude that their 
overall usefulness in coupling helicity determination will require very high integrated luminosities beyond the 5 ab$^{-1}$ employed in this 
survey unless the $W'$ is significantly lighter than our 12 TeV test example (by at least a few TeV). Clearly the shape of the $M_T$ distribution 
appears to be the lone observable employing `modest' statistics that will be available  to determine the $W'$ coupling helicity.

\section{Conclusions}

$Z'$ and $W'$ production are standard benchmarks for study of the physics capabilities of any new collider. In this paper, we have examined the 
production properties of these new particles at the FHC, in particular, with eye towards understanding its ability to explore their couplings to 
the usual SM fermions. This is a necessary first step in uncovering the underlying theory from which such new particles might spring. Of course 
the FHC is a long way from where we are now and one needs to make extrapolations on the theoretical side (\eg, pdfs, \etc) as well as make some 
guesses as to what objects might be well measured on the experimental side at such energies. For example, final states composed of purely 
hadronic objects or observables requiring b-tagging might be generally difficult to employ given the expected number of multiple interactions 
at 100 TeV with reasonable luminosities. Furthermore, it is likely that muon momenta in the multi-TeV range would be difficult to measure with 
high precision without enormous magnetic fields and/or bending radii.  Here, we have assumed that electron energies (and to some extent MET for 
the $W'$ case) can be as well measured by the FHC experiments as is presently done by the LHC experiments. Given these caveats, it is quite clear 
that the FHC will not only provide a huge step in the reach for these $Z'$ and $W'$ states but will allow for a sufficient number of electron-based 
observables to be examined so that we can learn something substantial about their couplings to the SM. If multiple interactions could be dealt with 
and muon-based observables can also be employed, these conclusions will only be strengthened. It is not likely, however, that such information 
would be obtainable for new gauge boson masses which lie within $\sim 5-10$ TeV of the corresponding discovery reaches without significant luminosity 
increases after the fact (due to simple statistical limitations) or through the application of observables we have not considered here. In performing 
these studies we have employed a very familiar and long-used set of models with which we can make direct comparisons to other previous experimental 
and theoretical analyses. However, it is important to recall that this set, though somewhat representative of a more general class of theories, does 
not in any way span the full range of possibilities that experimenters might encounter at the FHC. Further study of new gauge boson physics at the 
FHC is certainly needed. 

Clearly, much work will be required to understand what the full capabilities of the FHC might be. However,  it is clear that the FHC would provide a 
giant step forward in the search for BSM physics and hopefully its construction will be realized sometime in the future.

\section*{Acknowledgments}

The author would like to thank J.L. Hewett for discussions related to the work. He would also like to thank S. Godfrey for discussions about 
$Z'$ discovery reaches during the Snowmass meeting in Minneapolis. This work was supported by the Department of Energy, Contract DE-AC02-76SF00515.

\section*{Appendix}

It is sometimes useful to ask what would happen to the discovery reach for $Z'$ and $W'$ gauge bosons if their couplings differed from those in 
specified in a given model, particularly if they were weaker. To attempt to address this question here, we briefly consider the SSM $Z'/W'$ gauge 
bosons but with couplings that differ in overall strengths than those discussed above. Extrapolation to {\it larger} couplings is rather 
straightforward 
as in the mass ranges of interest the signals remain in a region where there is essentially no (non-fake) SM backgrounds. This is of course no 
longer true when we consider {\it weaker} couplings and correspondingly smaller masses. In particular, once a region of low mass with significant 
SM background is reached one quickly becomes dominated by the systematic errors associated with our knowledge of this background due to the large 
integrated luminosities that we consider. This is certainly a subject worthy of some detailed study by the future FHC experimental collaborations 
as this relies on more of the detailed properties of possible detectors than is within the scope of the present work. {\it If} we assume that these 
backgrounds can be controlled and understood at the same $\simeq 25\%$ level of uncertainty as claimed by ATLAS and CMS for the HL-LHC\cite{HL-LHC}, 
then we can get an estimate of what the discovery reach might be for these $Z'/W'$ states with weaker couplings. To get an idea of what might be 
possible at the FHC we will make this assumption below for the dilepton invariant mass region above a few TeV.

\begin{figure}[htbp]
\centerline{\includegraphics[width=5.0in,angle=90]{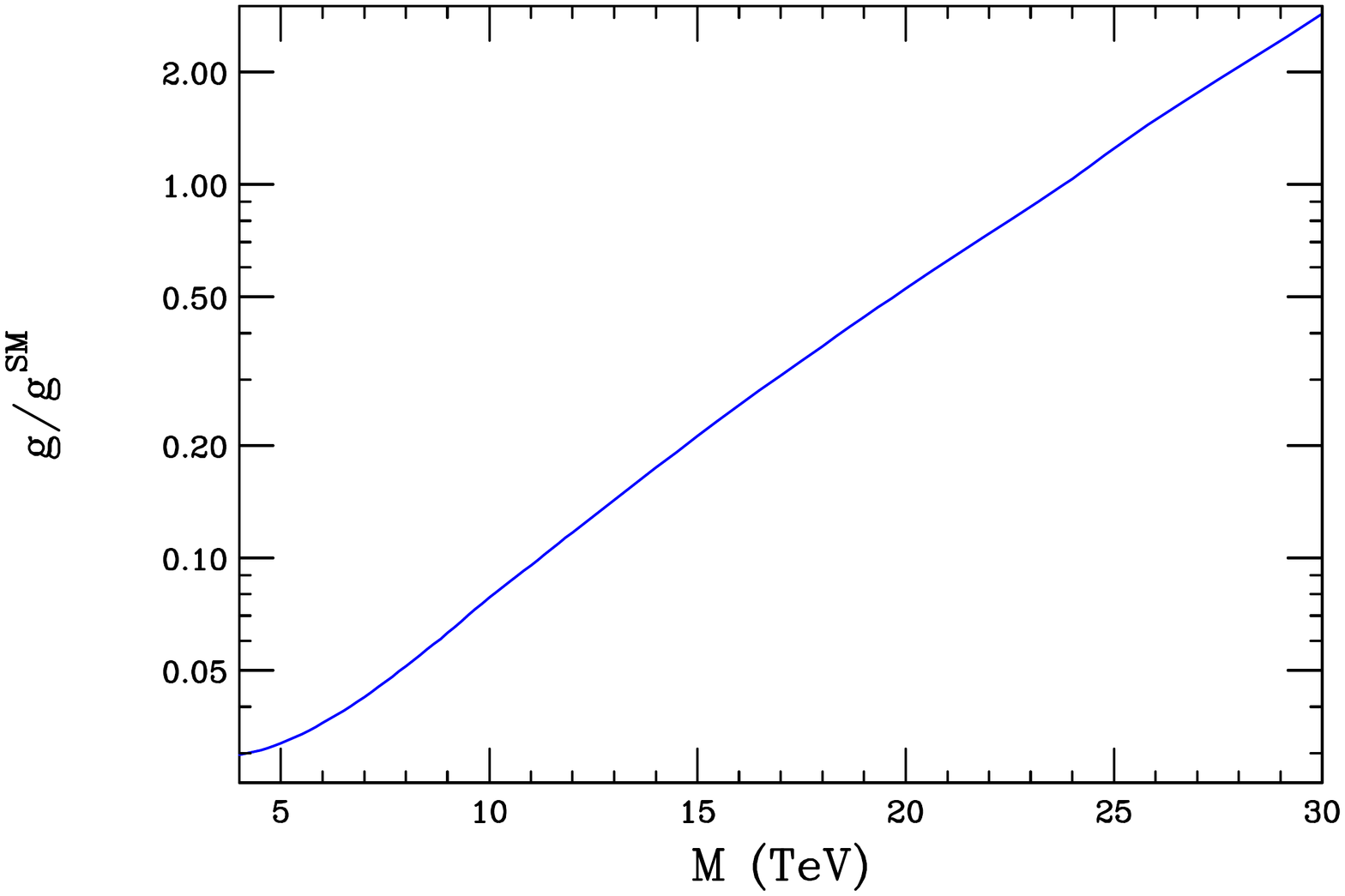}}
\vspace*{-4.0cm}
\centerline{\includegraphics[width=5.0in,angle=90]{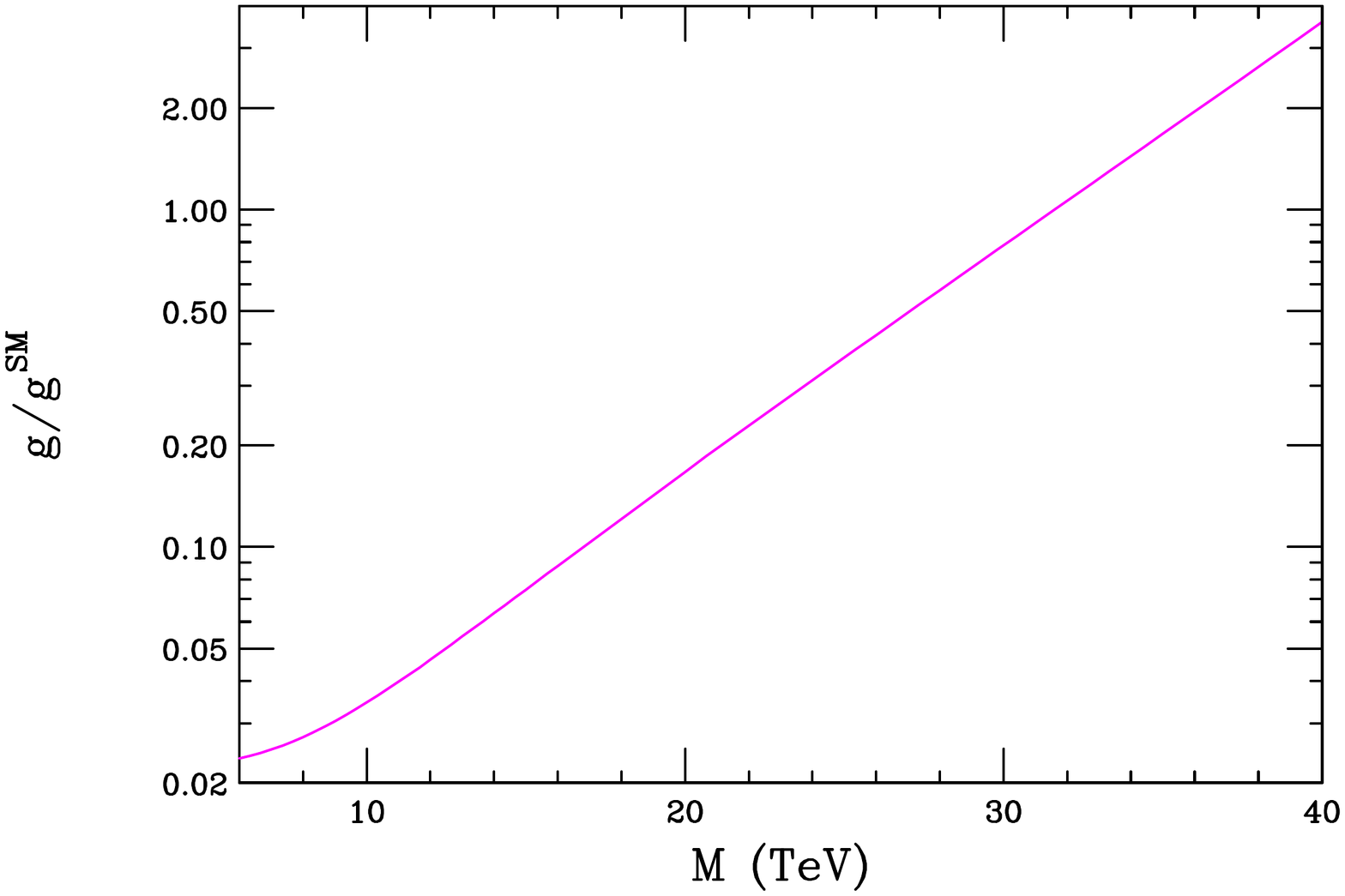}}
\vspace*{-1.50cm}
\caption{Discovery reach for SSM gauge bosons as a function of overall coupling strength assuming an integrated luminosity of 1 ab$^{-1}$: (top) 
$Z'$ and (bottom) $W'$.}
\label{fig16}
\end{figure}

Following this approach we obtain the results as shown in Fig.~\ref{fig16} for both the SSM $Z'$ and $W'$ with scaled couplings. In the $Z'$ case, 
we see that for $g/g_{SM}=0.1(0.05)$ the discovery reach for 1 ab$^{-1}$ is reduced to $\simeq 11(8)$ TeV. In the corresponding $W'$ case we see 
that the discovery reach is reduced to $\simeq 17(13)$ TeV. In both cases we note that at very low masses and small relative couplings the search 
reaches  `saturate' in the sense that smaller couplings cannot be probed with this level of systematic error given our rather simplified approach. 
This happens for masses below $\simeq 5$ TeV, with corresponding couplings less than $g/g_{SM}=0.025$, in the case of the $W'$ and at masses of 
$\simeq 4$ TeV and the corresponding coupling below $g/g_{SM}=0.02$ for the $Z'$. We emphasize to the reader that these results are only approximate 
but do give us some indication of what might be achievable at the FHC.

\end{document}